\documentclass[aip,chaos,amsmath,amssymb,reprint,preprintnumbers]{revtex4-1}

\usepackage{cmap}
\usepackage{indentfirst}
\usepackage{amssymb,amsmath,amsthm}
\usepackage{type1ec}
\usepackage[pdftex]{graphicx}
\usepackage{dcolumn}
\usepackage{bm}
\usepackage{color}
\newcommand{\trn}{^{\rm\scriptscriptstyle T}}
\newcommand{\e}{\rm e}

\renewcommand\epsilon{\varepsilon}
\renewcommand\le{\leqslant}
\renewcommand\ge{\geqslant}
\usepackage[
  a4paper, mag=1000, includefoot,
  left=3cm, right=1.5cm, top=2cm, bottom=2cm, headsep=1cm, footskip=1cm
]{geometry}

\newcounter{commentzaehler}

\begin{document}
\title{Synchronization in heterogeneous FitzHugh-Nagumo networks with hierarchical architecture} 
\author{S.A.~Plotnikov}
\affiliation{Saint Petersburg State University, St.~Petersburg, Russia}
\affiliation{ITMO University, St. Petersburg, Russia}
\affiliation{ Institute for Problems
of Mechanical Engineering, Russian Academy of Sciences,
Bolshoy~Ave,~61, Vasilievsky Ostrov,~St.~Petersburg, 199178~Russia }
\author{J.~Lehnert}
\affiliation{Institut f\"{u}r Theoretische Physik, TU Berlin,
Hardenbergstra{\ss}e 36, D-10623 Berlin, Germany}
\author{A.L.~Fradkov}
\affiliation{Saint Petersburg State University, St.~Petersburg, Russia}
\affiliation{ITMO University, St. Petersburg, Russia}
\affiliation{ Institute for Problems
of Mechanical Engineering, Russian Academy of Sciences,
Bolshoy~Ave,~61, Vasilievsky Ostrov,~St.~Petersburg, 199178~Russia }
\author{E.~Sch\"oll}
\affiliation{Institut f\"{u}r Theoretische Physik, TU Berlin,
Hardenbergstra{\ss}e 36, D-10623 Berlin, Germany}
\date{\today}

\begin{abstract}
We study synchronization in heterogeneous FitzHugh-Nagumo networks. It is well known that heterogeneities in the nodes
hinder synchronization when becoming too large. Here, we develop a
controller to counteract the impact of these heterogeneities. 
We first analyze the stability of the equilibrium point in a ring network of
heterogeneous nodes. We then derive a sufficient condition for synchronization in the absence
of control.
Based on these results we derive the controller providing 
synchronization for parameter values where synchronization without
control is absent. We demonstrate our results in networks with
different topologies. Particular attention is given to hierarchical (fractal) topologies, which are relevant for the architecture of the brain.

\end{abstract}

\maketitle

\section{Introduction}

Synchronization in systems of coupled oscillators has been investigated in various fields such as nonlinear dynamics, network science, and statistical physics and has applications in physiscs, biology, and technology \cite{BLE88,PIK01}. The examples of nontrivial collective dynamics in such ensembles include synchronous regimes in arrays of Josephson junctions \cite{WIE95} and lasers \cite{WIE90}, coordinated firing of cardiac pacemaker cells \cite{PES75}, synchronous emission of light pulses by a population of fireflies \cite{BUC68} and emission of chirps by a population of crickets \cite{WAL69}, and synchronization in ensembles of electrochemical oscillators \cite{KIS02}.

An important example of this class is related to the collective
dynamics of neuronal populations. Indeed, synchronization of
individual neurons is believed to play the crucial role in the
emergence of pathological rhythmic brain activity in Parkinson's
disease, essential tremor, and epilepsies; a detailed discussion of
this topic and numerous citations can be found in
Refs.~\onlinecite{TAS99,GOL01,MIL03}. Obviously, the development of
techniques for suppression of the undesired neural synchrony
constitutes an important clinical problem. Technically, this problem
can be solved by means of implantation of microelectrodes into the
impaired part of the brain with subsequent electric stimulation
through these electrodes \cite{CHK78,BEH91}. A fundamental
understanding of synchronization and its control can help to improve
the results of such a treatment by reducing at the same time its side effects.

Control of synchronization has so far focused on networks of identical nodes
\cite{ZHO08,LU09, LU12, SEL12, GUZ13, LEH14, SCH16}.
However, in realistic
networks the nodes will always be characterized by some diversity
meaning that the parameters of the different nodes are not identical
but drawn from a distribution.  It is well known that such
heterogeneities in the nodes can hinder or prevent synchronization and
that the coupling strength is a crucial parameter in this context
\cite{STR00,SUN09a}. Moreover, in most cases perfect synchronization -- in the sense
that the state of all nodes is identical at all times  -- is unfeasible in
the presence of  heterogeneities.

In order to grasp the complicated interaction of neurons in large
neural networks, those are often lumped into groups of neural
populations, each of which can be represented as an effective excitable
element that is mutually coupled to other elements
\cite{ROS04,ROS04a,POP05}. In this sense the simplest model which may
reveal features of interacting neurons consists of two or a few  coupled neural oscillators. Each of these can be represented by a simplified FitzHugh-Nagumo (FHN) system \cite{FIT61a,NAG62}. 

In the present paper we develop a 
control algorithm ensuring collective synchrony in heterogeneous
FitzHugh-Nagumo networks. Considerable attention is paid to studying
irregular coupling topologies; the motivation of this comes from
recent results in the area of neuroscience. Diffusion tensor magnetic
resonance imaging (DT-MRI) studies have revealed an intricate
architecture in the neuron interconnectivity of the human and
mammalian brain, which has already been used in simulations
\cite{VUK14}. The analysis of DT-MRI images (resolution of the order
of $0.5$ mm) has shown that the connectivity of the neuron axons
network represents a hierarchical geometry with fractal dimensions
varying between $2.3$ and $2.8$, depending on the local properties, on
the subject, and on the noise reduction threshold
\cite{KAT09,EXP11,PRO12}. The fractal connectivity dictates a hierarchical ordering in the
distribution of neurons which is essential for the fast and optimal
handling of information in the brain \cite{OME15}. To take this into account, we consider as an
application of our method a hierarchical
model which well represents these fractal properties of
the brain.

The rest of the paper is organized as follows:
Section~\ref{Sec:model_eq} introduces the model, while
Sec.~\ref{Sec:FHN_analys} analyzes the behavior of the FHN network: The
possible bifurcation scenarios are investigated for a unidirectional
ring topology, and the influence of the coupling strength on the
synchronizability of the network is discussed. In Sec.~\ref{Sec:analys_hier}, hierarchical
topologies are studied in dependence on different system
parameters. Section~\ref{Sec:control} develops the control algorithm and provides the simulation results
of controlling synchrony in different network topologies. Finally, we conclude with Sec.~\ref{Sec:concl}.

\section{Model equation}\label{Sec:model_eq}
The local dynamics of each node in the network is modeled by the
FitzHugh-Nagumo (FHN) differential equations \cite{FIT61a,NAG62}. The
FHN model is paradigmatic for excitable dynamics of type II, i.e., close to a Hopf
bifurcation \cite{LIN04}, a bifurcation which is not only
characteristic for neurons but also occurs in the context of other
systems ranging from electronic circuits \cite{HEI10} to
cardiovascular tissues \cite{WIN91,NAS04} and the climate system \cite{GAN02}. The
$i$th node of the network is described as follows:
\begin{equation}\label{m}
\begin{aligned}
\epsilon \dot{u_i}&=u_i-\frac{u_i^3}{3}-v_i+C\sum\limits_{j=1}^{N}G_{ij}[u_j(t-\tau)-u_i(t)], \\
\dot{v_i}&=u_i+a_i,\quad i=1,\dots,N,
\end{aligned}
\end{equation}
where $u_i$ and $v_i$ denote the activator and inhibitor variable of
the nodes $i=1,\dots,N$, respectively. $\tau$ is the delay, i.e., the
time the signal needs to propagate between node  $i$ and $j$. $\epsilon$ is a time-scale parameter and
typically small (here we will use $\epsilon=0.01$), i.e., $u_i$
is a fast variable, while $v_i$ changes slowly. The coupling matrix $\mathbf{G}=\{G_{ij}\}$
defines which nodes are connected to each other. We construct the
matrix $\mathbf{G}$ by setting the entry $G_{ij}$ equal to $1$~(or $0$) if
the $j$th node couples (or does not couple) into the $i$th node. The overall coupling strength is given by $C$.

In the uncoupled system ($C=0$), $a_i$ is a threshold parameter: For
$|a_i|>1$ the $i$th node is excitable, while for $|a_i|<1$
it exhibits self-sustained periodic firing. This is due to a
supercritical Hopf bifurcation at $|a_i|=1$ with a locally stable equilibrium
point for $|a_i|>1$ and a stable limit cycle for $|a_i|<1$. 
Here, we assume that the threshold parameters $a_i$ are chosen from the
interval $|a_i-\mu|<\sigma$ for some $\mu$.  For the analytical
considerations presented in Sec.~\ref{Sec:FHN_analys}, we do not make any further
assumption on the probability distribution of the $a_i$, while
for the simulation in Sec.~\ref{Sec:analys_hier} we use a Gaussian
distribution with mean  $\mu$ where we discard values of $a_i$ for which
$|a_i-\mu|<\sigma$  is not fulfilled. 
For $\mu=1$,  oscillatory and excitable nodes coexist.

\section{Analysis of FitzHugh-Nagumo network}\label{Sec:FHN_analys}
This section studies the behavior of a FHN network of heterogeneous
nodes in the absence of the controller.  First, we perform a linear stability
analysis of the equilibrium point to get insight into the possible
bifurcations. Second, we discuss for which  coupling strengths the
network synchronizes. Afterwards, we study the
impact of the coupling strength on the type of synchronization, i.e.,
whether synchronization takes place in the equilibrium or in an oscillatory
state.  

\subsection{Linear stability of the equilibrium point}\label{Sec:Lin_stab}

The linear stability analysis follows the approach suggested in
Refs.~\onlinecite{DAH08c,SCH08,PLO15}. Consider the unidirectional
ring of $N$ nodes described by the FHN equation with heterogeneous
threshold parameters, i.e., system \eqref{m} coupled by the following
adjacency matrix
\begin{equation}\label{eq1}
\mathbf{G} = \begin{pmatrix}
0 & 1 & 0 &\cdots & 0 \\
0 & 0 & 1 &\cdots & 0 \\        
\vdots & \vdots & \vdots &\ddots & \vdots \\
0 & 0 & 0 &\cdots & 1 \\
1 & 0 & 0 &\cdots & 0
\end{pmatrix}.
\end{equation}

The unique equilibrium point of the system \eqref{m} with coupling matrix \eqref{eq1} is given by $\mathbf{x}^*\equiv(u_1^*,v_1^*,\dots,u_N^*,v_N^*)\trn$, where $u_j^*=-a_j$, $v_j^*=-a_j+a_j^3/3+C(a_j-a_{(j+1)\bmod N})$ for $j=1,\dots,N$. Linearizing Eq.~\eqref{m} with matrix \eqref{eq1} around the equilibrium point $\mathbf{x}^*$ by setting $\mathbf{x}(t)=\mathbf{x}^*+\delta\mathbf{x}(t)$, one obtains

\begin{multline}\label{eq2}
\delta \mathbf{\dot x}=\frac{1}{\epsilon}\begin{pmatrix}
\xi_1 & -1 & 0 &0&0& \cdots&0& 0 \\
\epsilon & 0 & 0 &0&0& \cdots &0& 0 \\        
0 & 0 & \xi_2 & -1  &0&\cdots&0 &0 \\
0 & 0 & \epsilon & 0&0&\cdots&0 &0\\
0 & 0 & 0 & 0&\xi_3&\cdots&0 &0\\
\vdots &\vdots &\vdots&\vdots&\vdots&\ddots&\vdots&\vdots\\
0 & 0 &0 &0&0&\cdots&\xi_N&-1\\
0& 0 &0&0&0&\cdots&\epsilon&0
\end{pmatrix}\delta\mathbf{x}(t)\\
+\frac{1}{\epsilon}\begin{pmatrix}
0 & 0 & C & 0 & 0 & \cdots&0 &0 \\
0 & 0 & 0 & 0 & 0 & \cdots&0 & 0\\        
0 & 0 & 0 & 0 & C  &\cdots & 0& 0\\
0 & 0 & 0 & 0 & 0  &\cdots & 0 & 0\\
0 & 0 & 0 & 0 & 0  &\cdots & 0 & 0\\
\vdots &\vdots &\vdots&\vdots&\vdots&\ddots&\vdots&\vdots\\
C & 0 &0 & 0  & 0 & \cdots& 0 & 0\\
0 & 0 & 0 & 0  & 0& \cdots& 0 & 0
\end{pmatrix}\delta\mathbf{x}(t-\tau),
\end{multline}
where $\xi_j=1-a_j^2-C$ for $j=1,\dots,N$. The substitution
\begin{equation}\label{eq3}
\delta \mathbf{x}(t)=\e^{\lambda t}\mathbf{q},
\end{equation}
where $\mathbf{q}$ is an eigenvector of the Jacobian matrix, leads to the characteristic equation for the eigenvalue $\lambda$:
\begin{equation}\label{eq4}
\begin{vmatrix}
\xi_1-\epsilon\lambda & -1 & C\e^{-\lambda\tau} &0& \cdots& 0&0 \\
\epsilon & -\epsilon\lambda & 0 &0& \cdots & 0 &0\\    
0 & 0 & \xi_2-\epsilon\lambda& -1  &\cdots &0 &0\\
0 & 0 & \epsilon & -\epsilon\lambda&\cdots &0&0\\
\vdots &\vdots &\vdots&\vdots&\ddots&\vdots&\vdots\\
C\e^{-\lambda\tau} & 0&0&0 &\cdots&\xi_N-\epsilon\lambda&-1\\
0&0&0&0&\cdots&\epsilon&-\epsilon\lambda    
\end{vmatrix}=0.
\end{equation}

We can express the $(2N-1)$th row of the obtained matrix as a sum of two
rows rewriting Eq.~\eqref{eq4}  as
\begin{multline}\label{eq5}
\begin{vmatrix}
\xi_1-\epsilon\lambda & -1 & C\e^{-\lambda\tau} &0& \cdots& 0 & 0\\
\epsilon & -\epsilon\lambda & 0 &0& \cdots & 0& 0 \\    
0 & 0 & \xi_2-\epsilon\lambda& -1  &\cdots &0& 0 \\
0 & 0 & \epsilon & -\epsilon\lambda&\cdots &0& 0\\
\vdots &\vdots &\vdots&\vdots&\ddots&\vdots&\vdots\\
0 & 0&0&0&\cdots &\xi_{N}-\epsilon\lambda&-1\\
0&0&0&0&\cdots&\epsilon&-\epsilon\lambda    
\end{vmatrix}\\
+\begin{vmatrix}
\xi_1-\epsilon\lambda & -1 & C\e^{-\lambda\tau} &0& \cdots& 0 & 0\\
\epsilon & -\epsilon\lambda & 0 &0& \cdots & 0& 0 \\    
0 & 0 & \xi_2-\epsilon\lambda& -1  &\cdots &0 & 0\\
0 & 0 & \epsilon & -\epsilon\lambda&\cdots &0& 0\\
\vdots &\vdots &\vdots&\vdots&\ddots&\vdots&\vdots\\
C\e^{-\lambda\tau} & 0&0&0&\cdots &0&0\\
0&0&0&0&\cdots&\epsilon&-\epsilon\lambda    
\end{vmatrix}=0.
\end{multline}

Denote by $I_1$ the first matrix in the Eq.~\eqref{eq5} and by $I_2$
the second one, respectively. Firstly, we consider the determinant of
matrix $I_1$. We express the first row of $I_1$ as the sum of the rows
$(\xi_1-\epsilon\lambda,  -1,  0, 0, \ldots, 0, 0)$
and $(0, 0,  C\e^{-\lambda\tau},0, \ldots, 0, 0)$. In this way,
$I_1$ is written as the sum of two determinants. The first of these
two determinants is Laplace expanded
starting with its first row. As a result, we can rewrite $I_1$ as
\begin{multline}\label{eq6}
\det I_1 =\epsilon(1-\xi_1\lambda+\epsilon\lambda^2)\\
\times\begin{vmatrix}
\xi_2-\epsilon\lambda & -1 & C\e^{-\lambda\tau} &0& \cdots& 0 & 0\\
\epsilon & -\epsilon\lambda & 0 &0& \cdots & 0& 0 \\    
0 & 0 & \xi_3-\epsilon\lambda& -1 &\cdots &0& 0 \\
0 & 0 & \epsilon & -\epsilon\lambda&\cdots &0& 0\\
\vdots &\vdots &\vdots&\vdots&\ddots&\vdots&\vdots\\
0 & 0&0&0&\cdots &\xi_{N}-\epsilon\lambda&-1\\
0&0&0&0&\cdots&\epsilon&-\epsilon\lambda    
\end{vmatrix}\\
+\begin{vmatrix}
0 & 0 & C\e^{-\lambda\tau} &0& \cdots& 0 & 0\\
\epsilon & -\epsilon\lambda & 0 &0& \cdots & 0 & 0\\    
0 & 0 & \xi_2-\epsilon\lambda& -1  &\cdots &0 & 0\\
0 & 0 & \epsilon & -\epsilon\lambda&\cdots &0& 0\\
\vdots &\vdots &\vdots&\vdots&\ddots&\vdots&\vdots\\
0 & 0&0&0&\cdots &\xi_{N}-\epsilon\lambda&-1\\
0&0&0&0&\cdots&\epsilon&-\epsilon\lambda    
\end{vmatrix}.
\end{multline}
Evidently, the determinant of the second matrix in the sum \eqref{eq6} equals zero. By repeating the same procedure, we derive
\begin{equation}\label{eq7}
\det I_1 = \epsilon^N\prod\limits_{j=1}^{N}(1-\xi_j\lambda+\epsilon\lambda^2).
\end{equation}

Now let us  consider the determinant of matrix $I_2$.  Again, we use a
Laplace expansion; this time starting with the $(2N-1)$th row we get
\begin{multline}\label{eq8}
\det I_2 = C\e^{-\lambda\tau}\\
\times\begin{vmatrix}
 -1 & C\e^{-\lambda\tau} &0&0& \cdots& 0& 0 \\
 -\epsilon\lambda & 0 &0&0& \cdots & 0& 0 \\    
 0 & \xi_2-\epsilon\lambda& -1 &C\e^{-\lambda\tau} &\cdots &0 & 0\\
 0 & \epsilon & -\epsilon\lambda&0&\cdots &0& 0\\
\vdots &\vdots&\vdots&\vdots&\ddots&\vdots&\vdots\\
0&0&0&0&\cdots&0&0\\
0&0&0&0&\cdots&\epsilon&-\epsilon\lambda    
\end{vmatrix}.
\end{multline}
This matrix has dimension $(2N-1)\times(2N-1)$ and can be expanded by the last column:
\begin{multline}\label{eq9}
\det I_2 = -\epsilon\lambda C\e^{-\lambda\tau}\\
\times\begin{vmatrix}
 -1 & C\e^{-\lambda\tau} &0&0& \cdots& 0& 0  \\
 -\epsilon\lambda & 0 &0&0&\cdots & 0& 0  \\    
 0 & \xi_2-\epsilon\lambda& -1 &C\e^{-\lambda\tau} &\cdots &0& 0  \\
 0 & \epsilon & -\epsilon\lambda&0&\cdots &0& 0 \\
\vdots &\vdots&\vdots&\vdots&\ddots&\vdots&\vdots\\
0&0&0&0&\cdots& -1 & C\e^{-\lambda\tau} \\
0&0&0&0&\cdots&-\epsilon\lambda&0\\  
\end{vmatrix}.
\end{multline}

We continue expanding by the last column and finally obtain
\begin{equation}\label{eq10}
\det I_2 =-\epsilon^NC^N\lambda^N\e^{-\lambda\tau N}.
\end{equation}
Using Eqs.~\eqref{eq7} and \eqref{eq10} we obtain the characteristic equation for system \eqref{eq2}:
\begin{equation}\label{eq11}
\prod\limits_{j=1}^{N}(1-\xi_j\lambda+\epsilon\lambda^2)-(C\lambda\e^{-\lambda\tau})^N=0.
\end{equation}

We can neglect $\epsilon$ since $\epsilon\ll1$ holds, i.e., $\epsilon\approx0$ in the following.  Substituting $\lambda=i \omega$, which holds at the stability boundary given by a Hopf bifurcation, into the Eq.~\eqref{eq11} yields
\begin{equation}\label{eq12}
\prod\limits_{j=1}^{N}(1-\xi_j i\omega)=(C i\omega\e^{-i\omega\tau })^N.
\end{equation}

We take the squared absolute value of both sides of Eq.~\eqref{eq12}
leading to
\begin{equation}\label{eq13}
\prod\limits_{j=1}^{N}(1+\xi_j^2\omega^2)=(C \omega)^{2N}.
\end{equation}
Equation~\eqref{eq13} can be expressed as
\begin{equation}\label{eq14}
\left(\prod\limits_{j=1}^{N}\xi_j^2-C^{2N}\right)\omega^{2N}+\mathbf{P}(\omega^2)=0,
\end{equation}
where $\mathbf{P}(\omega^2)$ is the polynomial of $(N-1)$th degree and with positive coefficients. For $\prod_{j=1}^{N}\xi_j^2>C^{2N}$ Eq.~\eqref{eq14} and hence  Eq.~\eqref{eq12} has no solution for real valued $\omega$ and, thus, no Hopf bifurcation will take place. Taking the square root of this inequality and resubstituting $\xi_j=1-a_j^2-C$ yields
\begin{equation}\label{eq15}
\left|\prod\limits_{j=1}^{N}(1-C-a_j^2)\right|>|C|^{N}.
\end{equation}
This inequality defines the values of threshold parameters $a_j$ where
a Hopf bifurcation is impossible, i.e., the equilibrium point is
stable. This result is a generalization of the inequality obtained in
Ref.~\onlinecite{PLO15} for an unidirectional ring topology. For
sufficiently large values of the coupling strength and in the presence
of a large number of excitable nodes, i.e., $a_i>1$, the inequality
\eqref{eq15} is fulfilled, hence the whole network is in an excitable
state. Analytical treatment of this problem for other network
topologies is nontrivial. However, we expect qualitatively similar
results for other network topologies, for example with symmetric
coupling matrices, as will be discussed in
Sec.~\ref{Sec:coupl_strength_analys} and checked by the simulation in
Sec.~\ref{Sec:analys_hier}.

\subsection{Synchronization analysis}\label{Sec:synch_analys}

In this section, we study the conditions under which network~\eqref{m}
synchronizes. We will show that the coupling strength is the crucial
parameter determining the synchronizability of the network. Here, we will focus on the case of $\tau=0$, i.e., a
coupling without delay. This allows for an analytic
treatment of the problem. The simulations in Sec.~\ref{Sec:analys_hier} will show
that the results obtained for $\tau=0$ hold in very good approximation
for $\tau>0$ though the delay might influence the transient behavior.

Consider the FHN network \eqref{m} with heterogeneous threshold
parameters but without delay, i.e., $\tau=0$,
\begin{equation}\label{f1}
\begin{aligned}
\epsilon \dot u_i&=u_i-\frac{u_i^3}{3}-v_i+C\sum\limits_{j=1}^{N}G_{ij}[u_j-u_i],\\
\dot v_i&=u_i+a_i.
\end{aligned}
\end{equation}
We assume that the connectivity graph $\Gamma$ of network \eqref{f1} is connected
and undirected, i.e., its adjacency matrix $\mathbf{G}$ is symmetric
and has no zero rows. By averaging over all nodes we obtain an
averaged trajectory described by 
\begin{equation}\label{f2}
\begin{aligned}
\epsilon \dot {\bar u}&=\bar u-\psi(u_1,\dots,u_N)-\bar v,\\
\dot {\bar v}&=\bar u+\bar a, 
\end{aligned}
\end{equation}
where $\bar u = \frac{1}{N}\sum_{j=1}^{N}u_j$, $\bar v =
 \frac{1}{N}\sum_{j=1}^{N}v_j$, $\bar a =  \frac{1}{N}\sum_{j=1}^{N}a_j$, and
$\psi(u_1,\dots,u_N) =  \frac{1}{3N}\sum_{j=1}^{N}u_j^3$. It will be shown that
this averaged trajectory  approximates well the synchronized behavior. 

In the following, we want to investigate how synchrony spreads in a
network. For this purpose we consider a network where all but one node
(in the following: node $i$)
are in synchrony, and evaluate the conditions under which this node
will synchronize with the other nodes. This approximates the
situation where a population of neurons is affected by pathological
synchronization during epileptic seizure and the question arises under
which condition the synchrony spreads.

To this end, let us introduce a \textit{leader} system:
\begin{equation}\label{f3}
\begin{aligned}
\epsilon \dot u_L&= u_L-\frac{u_L^3}{3}-v_L,\\
\dot v_L&=u_L+\bar a, 
\end{aligned}
\end{equation}
which describes a  FHN system without coupling and a threshold
equal to the mean  $\bar a=  \frac{1}{N}\sum_{j=1}^{N}a_j$. The dynamics of the \textit{leader} system
approximates well the dynamics of the synchronized network \eqref{f1}
which can easily be seen by comparing Eq.~\eqref{f3} with  Eq.~\eqref{f2}.

Let us now consider the $i$th node of the network \eqref{f1}, i.e.,
the node which is not yet synchronized with the rest of the  network. Assume that
node $i$ has $n_i$ connections with its neighbors.  
Note that in this case the coupling to node $i$ can be approximated by
$\sum_{j=1}^NG_{ij}u_j\sim n_iu_L$. Keeping
this in mind, we subtract the first row of Eq.~\eqref{f3} from the
first one of  Eq.~\eqref{f1}, and the second one from the second one,
respectively.  In addition we make the following substitution
\begin{equation}\label{f4}
\begin{aligned}
\delta_{u}(t)&=u_i(t)-u_L(t)+a_i-\bar a, \\ \delta_{v}(t)&=v_i(t)-v_L(t)+d,
\end{aligned}
\end{equation}
where $d=(1-Cn_i)(a_i-\bar a)$ is a constant, and finally obtain 
\begin{equation}\label{f5}
\begin{aligned}
\epsilon \dot\delta_{u}&=(1-Cn_i)\delta_{u}-\delta_{v}-(\delta_{u}-a_i+\bar a)\phi, \\
\dot\delta_{v}&=\delta_{u},
\end{aligned}
\end{equation}
where $\phi(t)=\frac{1}{3}(u_i(t)^2+u_i(t)u_L(t)+u_L(t)^2)$, $\phi(t)\ge0~\forall t$ is a nonnegative function. 

We now introduce the following Lyapunov function
\begin{equation}\label{f6}
V(t,\mathbf{\Delta}(t))=\frac{\epsilon\delta_{u}^2(t)}{2}+\frac{\delta_{v}^2(t)}{2},
\end{equation}
where $\mathbf{\Delta}=(\delta_{u},\delta_{v})$, and find its
derivative according to Eq.~\eqref{f5}
\begin{multline}\label{f7}
\dot
V(t,\mathbf{\Delta}(t))=\delta_{u}[(1-Cn_i)\delta_{u}-\delta_{v}-(\delta_{u}-a_i+\bar
a)\phi] \\
+\delta_{v}\delta_{u}=(1-Cn_i)\delta_{u}^2-(\delta_{u}^2-(a_i-\bar a)\delta_{u})\phi.
\end{multline}

The first term in Eq.~\eqref{f7} is negative for $C>1/n_i$. With
$\phi(t)\ge0~\forall t$ we note that the second term in Eq.~\eqref{f7} is
nonpositive for $|\delta_{u}|>2\sigma$, since $|a_i-\mu|<\sigma$ and therefore $|a_i-\bar a|<2\sigma$.

In conclusion, the Lyapunov function derivative is negative and the Lyapunov function \eqref{f6} decreases.
Thus, we obtain the inequality $C>1/n_i$ as a  sufficient condition that the $i$th neuron
synchronizes with the other nodes with a  level of precision
equal to  $2\sigma$, i.e., $|u_i(t)-u_L(t)+a_i-\bar a |<2\sigma$ holds for $t>t_c$, where $t_c$ is the time
transients need to decay.

In the case that more than one node is not synchronized with the rest
of the network, an analytic
treatment is difficult in the presence of heterogeneous threshold
parameters.  However, the previous results in this Section suggest that in this case
the network synchronizes if $C>1/n_{\text{min}}$ where
$n_{\text{min}}$ is the minimal degree of the network, i.e., $n_{\text{min}}\le n_j$,
$j=1, \ldots, N$ and $n_j$ denotes the degree of the $j$th
node. In the synchronized state, the activator and
inhibitor variables fulfill $u_j-u_L \approx a_j-\bar
a$ and $v_j-v_L \approx (1-Cn_j)(a_j-\bar a)$, respectively.

If the inequality $C>1/n_{\text{min}}$ does not hold, the network might
desynchronize in which case control is needed to enforce a synchronized state.  A
control scheme will be discussed in Sec.~\ref{Sec:control}.

\subsection{Influence of  the coupling strength
  on the type of synchronization}\label{Sec:coupl_strength_analys}

In the last subsection, we have discussed the conditions under which the
network synchronizes  but have not specified in which state -- oscillatory
or excitatory -- the synchronization takes place.
Here, we investigate the influence of the coupling strength on the
type of synchronized dynamics.

 To this end  let us consider the $i$th neuron in network \eqref{m} and
 rewrite it as
\begin{equation}\label{f11}
\begin{aligned}
\epsilon \dot u_i&=u_i-\frac{u_i^3}{3}-v_i,\\
\dot v_i&=u_i+a_i-C\sum\limits_{j=1}^{N}G_{ij}[\dot u_j(t-\tau)-\dot u_i(t)], 
\end{aligned}
\end{equation}
where we use the following substitution  $v_i \rightarrow v_i-C\sum_{j=1}^{N}G_{ij}[u_j(t-\tau)- u_i(t)]$.

The behavior of the i$th$ node without coupling depends on the
threshold parameter $a_i$. From Eq.~\eqref{f11} it follows that the
coupling effectively acts as a shift of the threshold. If the $j$th
node is excitable, then the derivative of its activator equals
zero. Therefore, only nodes in the oscillatory state influence the
effective threshold $a_i-C\sum\limits_{j=1}^{N}G_{ij}[\dot u_j(t-\tau)-\dot u_i(t)]$.

Suppose that $a_i>1$, i.e., the i$th$ node is excitable. If the
coupling strength $C$ is too small, then the i$th$ node remains in
the excitable regime. Increasing the coupling strength forces the  $i$th node to exhibit self-sustained
periodic firing. If we now further increase the coupling strength,
the absolute value of the term $|a_i-C\sum_{j=1}^{N}G_{ij}[\dot u_j(t-\tau)-\dot u_i(t)]|$ becomes larger
than $1$ for large time intervals meaning that the $i$th node stops to oscillate.
Thus, if the coupling is too high, the whole system is
excitable. To summarize the results of this section, if we want to synchronize the FHN network in the
oscillatory regime, we should choose the coupling sufficiently large
for synchronization, however, not so huge that we leave the
oscillatory regime and synchronize the network in the equilibrium point.

\section{Analysis of hierarchical network topology}\label{Sec:analys_hier}

In this section, we consider a network with specific hierarchical
architecture and study numerically how the network synchronization depends on different
system parameters as the coupling strength $C$, the delay $\tau$,
the variance of threshold parameters $\sigma^2$, and the topology. For the
simulations we use the normally distributed threshold parameters $a_i$
with mean $\mu=1$ and variance $\sigma^2>0$ meaning that
oscillatory and excitable nodes coexist. We restrict
ourselves to  threshold parameters $a_i$ from the
interval $|a_i-\mu|<\sigma$. The study of
different hierarchical architectures in the neuron connectivity is
motivated by MRI results of the brain structure which show that the
neuron axons networks spans the brain area fractally and not
homogeneously \cite{KAT09,EXP11,KAT12, OME15}. In the rest of this
paper the word ``fractal'' will be employed to denote
hierarchical structures of finite order $n$, since the human brain
has finite size and does not cover all orders. This is in contrast to
the exact definition of a fractal set where  $n\to\infty$.

Simple hierarchical structures can be constructed using the classic
Cantor fractal construction process on a ring network \cite{MAN83, FED88}. Using the
iterative bottom-up procedure to construct the Cantor set, we create a
symbolic sequence consisting of $0$ and
$1$
hierarchically nested . Starting with a base pattern $B$ containing $b$ symbols ($0$ or
$1$) we iterate it $n$ times and obtain a system of size
$N=b^n$. During each iteration step the symbols $1$ and $0$ are
replaced by the base pattern $B$ and a series of $b$ zeros,
respectively. Thus, we get the string $G_1$ of size $N=b^n$. The
resulting string $G_1$ defines the connections of all neurons to
the first node.  We now add a $0$ before the first symbol of $G_1$, i.e., the
first node has no connection with itself. In this extended string, a
$1$ at position $i$, $i=2,\ldots,  b^n+1$, represents a link between  the  $i$th  node
and the first one, while a $0$ means that these two nodes do not
couple. To construct the coupling matrix $\mathbf{G}$, we regard
$G_1$ as the first row in the matrix $\mathbf{G}$. Each of the
following rows are obtained by shifting the preceding
row cyclically by one element resulting in a $(b^n+1)\times (b^n+1)$ circulant matrix, i.e.,
$G_{ij}=G_{1,(j-i+N)\bmod N}$, $i,j=1,\ldots,  b^n+1$. 
The matrix constructed in this way contains a hierarchical
distribution of gaps with a variety of sizes. The following equation
describes the first three rows of a coupling matrix with hierarchical topology with the base $B=[1~0~1]$, $b=3$, $n=2$:
\begin{equation}\label{f12}
\mathbf{G} = \begin{pmatrix}
0 & 1 & 0 &1 &0 & 0 &0&1&0&1 \\
1&0 & 1 & 0 &1 &0 & 0 &0&1&0\\
0&1&0 & 1 & 0 &1 &0 & 0 &0&1\\
\cdots &\cdots &\cdots & \cdots &\cdots&\cdots&\cdots&\cdots&\cdots&\cdots 
\end{pmatrix}.
\end{equation}

The number of times the symbol $1$ appears in the base, denoted by
$c_1$, defines formally the fractal dimension $d_f$ of the structure,
as $d_f=\ln c_1/\ln b$. This measure $d_f$ describes perfectly the
fractal structure when the number of iterations $n\to\infty$. Note that for symmetric base patterns the adjacency matrices of networks
constructed according to this procedure are always symmetric and, thus, the results
obtained in Sec.~\ref{Sec:FHN_analys} hold for $\tau=0$ and in good
approximation for $\tau>0$.  

\begin{figure}
\begin{minipage}[h]{1\linewidth}
\center{\includegraphics[width=1\linewidth]{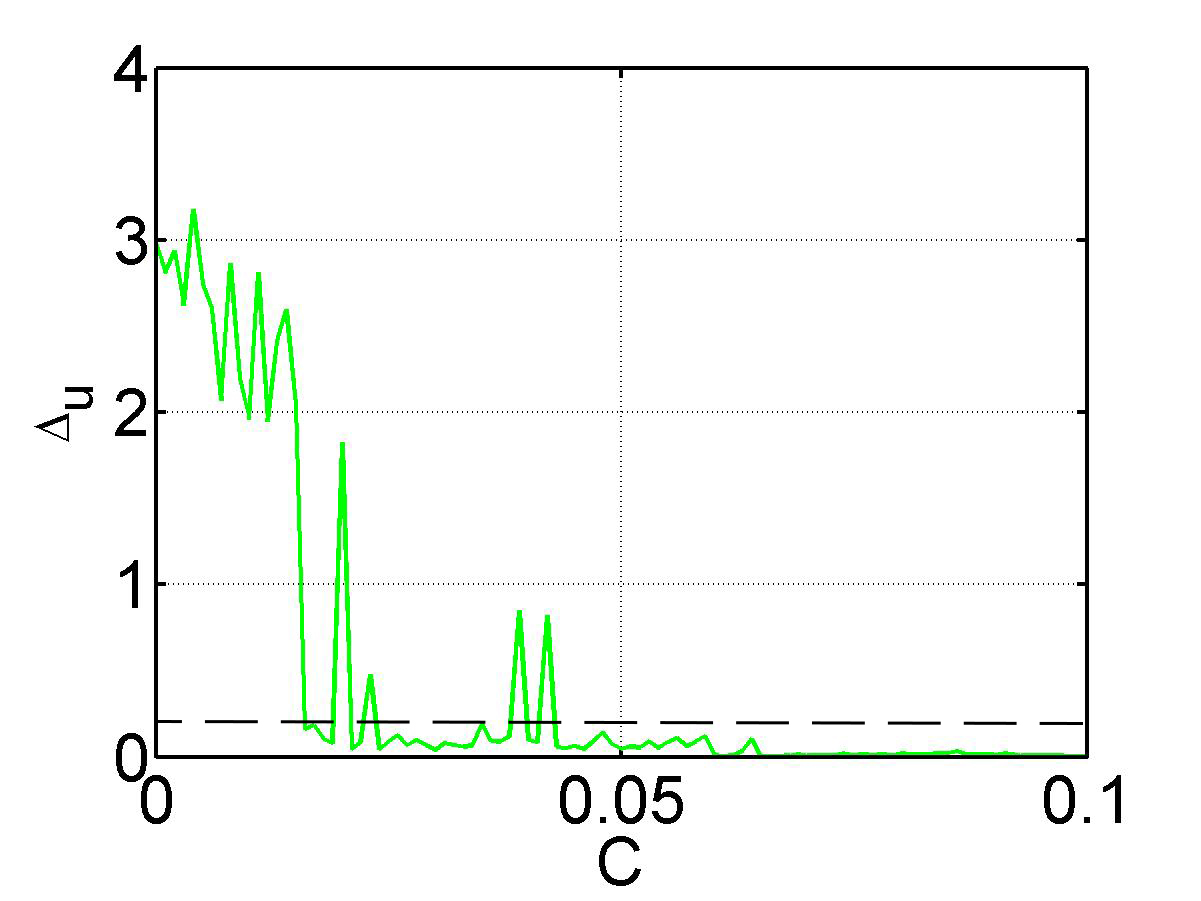}}
\end{minipage}
\caption{Dependence of the activator synchronization index $\Delta_u$
  on the coupling strength $C$ for the FitzHugh-Nagumo network given
  by Eq.~\eqref{m} with Cantor-type hierarchical topology. The black dashed line indicates the synchrony condition $\Delta_u<2 \sigma$. Parameters:
  $B=[1~0~1]$, $b=3$, $n=4$, $N=82$, $\tau=1.5$, $\epsilon=0.01$. The
  threshold parameters $a_i$ are normally distributed with mean $\mu=1$ and standard deviation $\sigma=0.1$, truncated at $a_i=\mu\pm\sigma$. Initial conditions: $u_i(t)=0$, $v_i(t)=0$, $i=1,\dots,N$,  for $t\in[-\tau,0]$.}
\label{figC}
\end{figure}

\begin{figure}
\begin{minipage}[h]{1\linewidth}
\center{\includegraphics[width=1\linewidth]{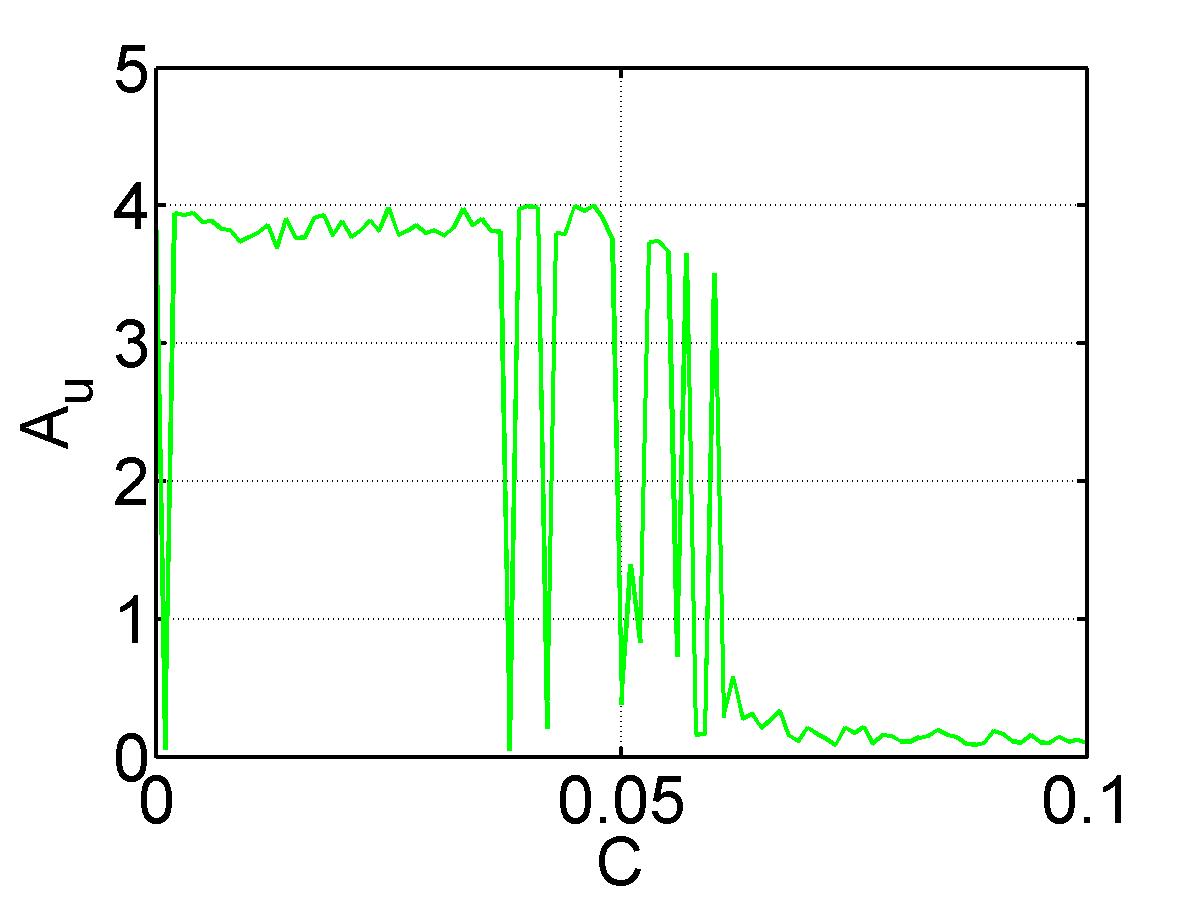}}
\end{minipage}
\caption{Dependence of the activator oscillation amplitude $A_u$ of the first node on the coupling strength $C$ for the FitzHugh-Nagumo network given by Eq.~\eqref{m} with Cantor-type hierarchical topology. Parameters and initial conditions as in Fig.~\ref{figC}.}
\label{figAmp}
\end{figure}

\begin{figure}
\begin{minipage}[h]{1\linewidth}
\center{\includegraphics[width=1\linewidth]{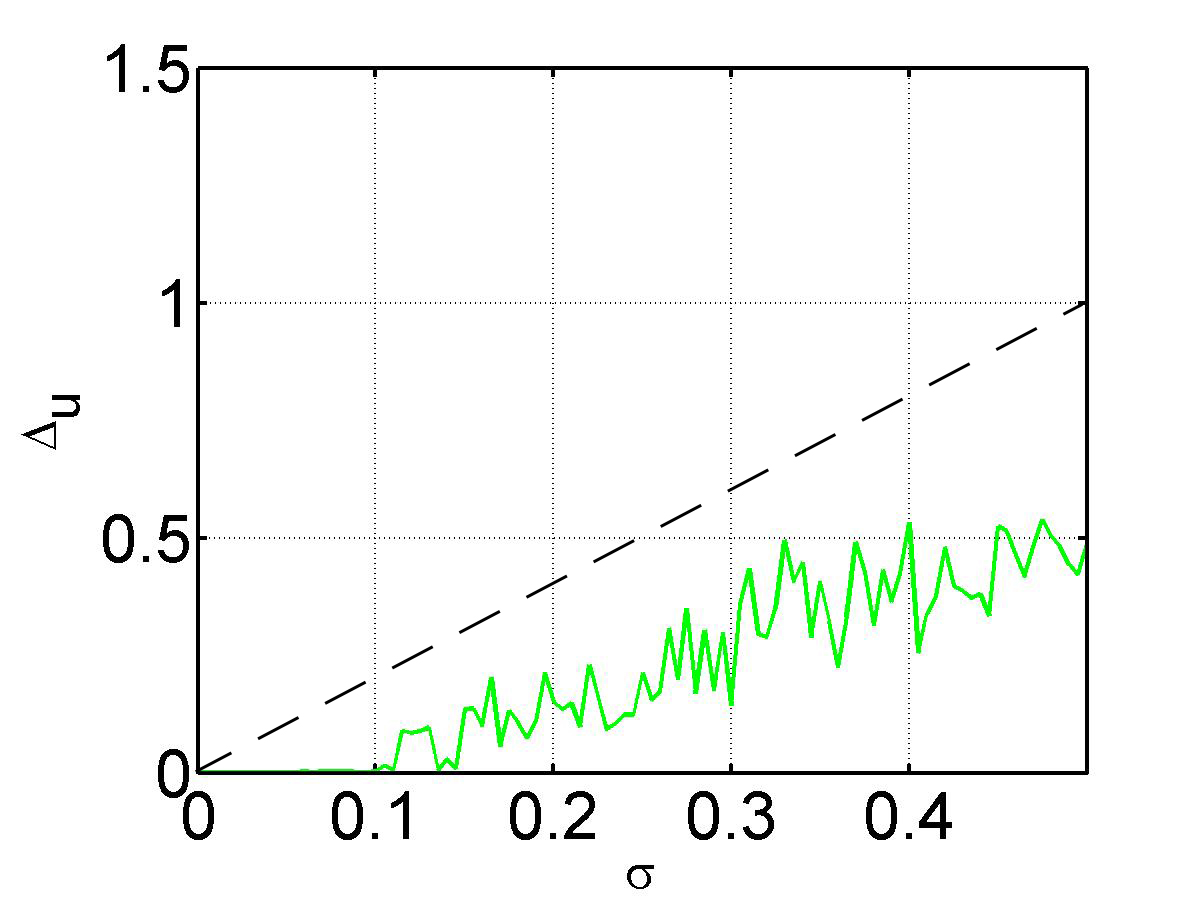}}
\end{minipage}
\caption{Dependence of the activator synchronization index $\Delta_u$ on the standard deviation $\sigma$ of threshold parameters $a_i$ for the FitzHugh-Nagumo network given by Eq.~\eqref{m} with Cantor-type hierarchical topology. The black dashed line indicates the synchrony condition $\Delta_u<2 \sigma$. Parameters: $C=0.065$. Other parameters and initial conditions as in Fig.~\ref{figC}.}
\label{figSig}
\end{figure}

\begin{figure}
\begin{minipage}[h]{1\linewidth}
\center{\includegraphics[width=1\linewidth]{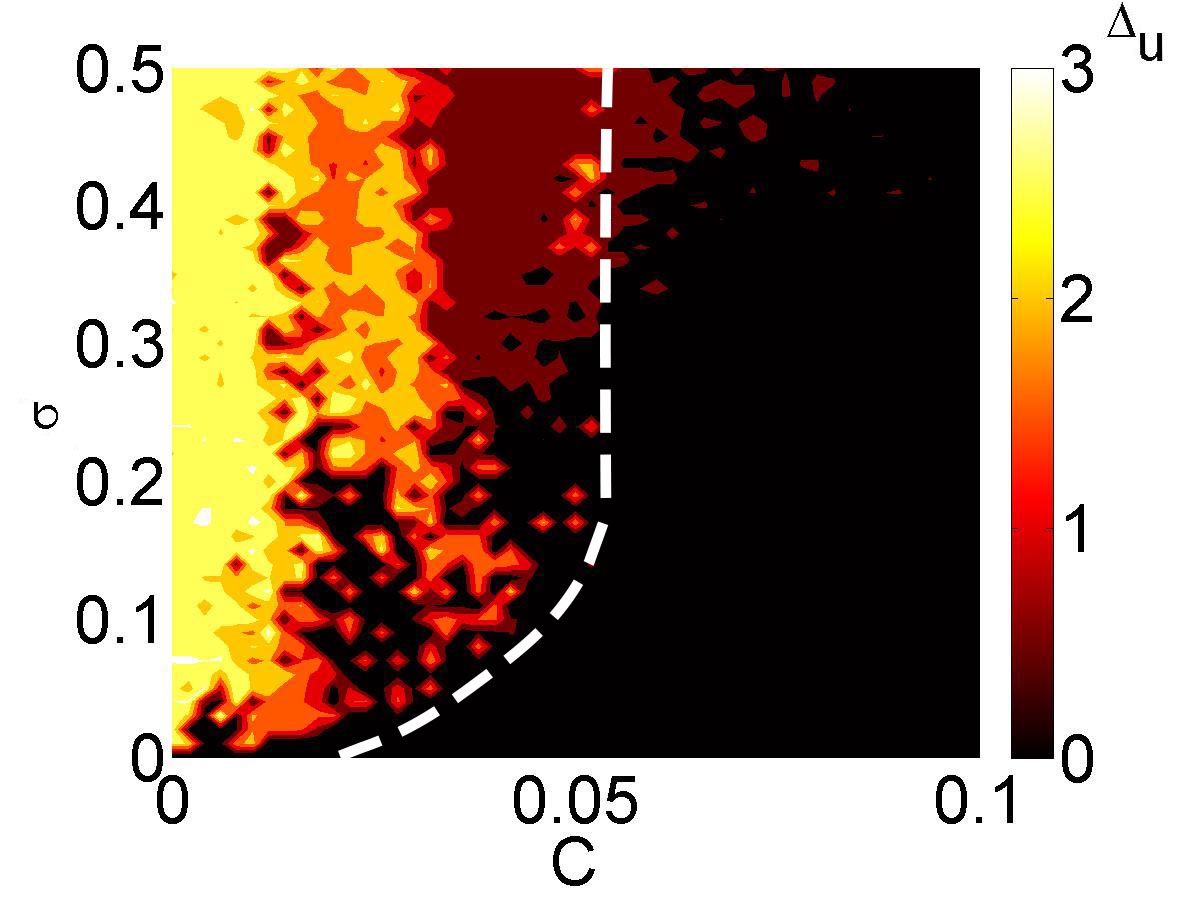}}
\end{minipage}
\caption{Dependence of the activator synchronization index $\Delta_u$ on the coupling strength $C$ and the standard deviation $\sigma$ of threshold parameters $a_i$ for the FitzHugh-Nagumo network given by Eq.~\eqref{m} with Cantor-type hierarchical topology. The white dashed curve  indicates the synchrony boundary $\Delta_u=2 \sigma$. To the right of this curve synchronization holds. Parameters and initial conditions as in Fig.~\ref{figC}.}
\label{figCSig}
\end{figure}

The simulations were carried out in the {\it Matlab} environment using the function $\it dde23$.

In the current study we use as base pattern $B=[1~0~1]$, i.e.,
$b=3$.  We perform $n=4$ iterations yielding a network of size $N=b^4+1=82$ nodes. For this base the value of $c_1$
(number of times the symbol $1$ is encountered within the base $B$)
equals $2$, i.e., the connectivity matrix has a fractal dimension of 
$d_f=\ln2/\ln3=0.6309$. After $n=4$ iterations, each node is connected
to  $n_i=c_1^n=2^4=16$ other nodes, while no connection to the
remaining $b^n+1-n_i=66$ elements exists. From the results obtained in
Sec.~\ref{Sec:synch_analys}, we expect the network to synchronize at $C>1/n_i=0.0625$.

We now consider a network with a delay $\tau=1.5$ and a standard
deviation of $\sigma=0.1$ for the threshold parameters $a_i$. We change
the coupling strength $C$ from $0$ to $0.1$ and check whether the network
synchronizes. To characterize the degree of synchrony, we introduce
the activator synchronization index
$\Delta_u=\max_{i=1}^{N}|u_i(t^*)-\bar u(t^*)+a_i-\bar a|$, where we define synchrony by the condition $\Delta_u<2\sigma$. Here we
use $t^*=30\ge t_c$, where $t_c$ is the characteristic transient time, to analyze the network after the decay of all
transients. The results of the simulation are presented in
Fig.~\ref{figC}, where the synchronization index vs. the coupling
strength is shown. One can see that there is a critical coupling strength $C^*\approx0.045$,
such that for $C>C^*$ synchronization holds with some level of
precision. To distinguish whether the network is in an oscillatory or excitable
state we  introduce the activator oscillation amplitude of the first
node $A_u$ as the difference between the maximum and minimum values of
the activator after the transient. Figure~\ref{figAmp} shows the
dependence of the activator oscillation amplitude of the first node
$A_u$ on the coupling strength $C$ showing that for $C>0.06$ the
$1$st node is in the excitable state. Thus, by combining the results of
the simulations shown in Figs.~\ref{figC} and~\ref{figAmp} we conclude
that for $0.045<C<0.06$ we obtain synchronization in an oscillatory state.
The results of the simulations confirm the idea discussed in Sec.~\ref{Sec:coupl_strength_analys}.

Now we fix the delay $\tau=1.5$ and the coupling strength $C=0.065$ and
change the standard deviation $\sigma$ from $0$ to $0.5$. The results
of the simulation are shown in Fig.~\ref{figSig}. As anticipated, the synchronization error increases
as $\sigma$ is enlarged. For all $\sigma$ the activator synchronization
index is below $2\sigma$ marked by a black dashed line. This is the level of precision of
synchronization predicted in Sec.~\ref{Sec:synch_analys}. Note that for
$\sigma<0.1$ the activator synchronization index $\Delta_u$ is close
to zero. To raise the level of precision for $\sigma>0.1$, we can increase the coupling strength $C$. 

Figure~\ref{figCSig} shows the value of the activator synchronization
index $\Delta_u$ in the $C$-$\sigma$ plane. The black area corresponds
to parameter values where the network synchronizes. The value
$C>1/n_\text{min}=0.0625$ approximates well the threshold between
synchronization and desynchronization as predicted in
Sec.~\ref{Sec:synch_analys}. For increasing $\sigma$, $\Delta_u$
increases showing that synchronization with less precision takes place.

\begin{figure}
\begin{minipage}[h]{1\linewidth}
\center{\includegraphics[width=1\linewidth]{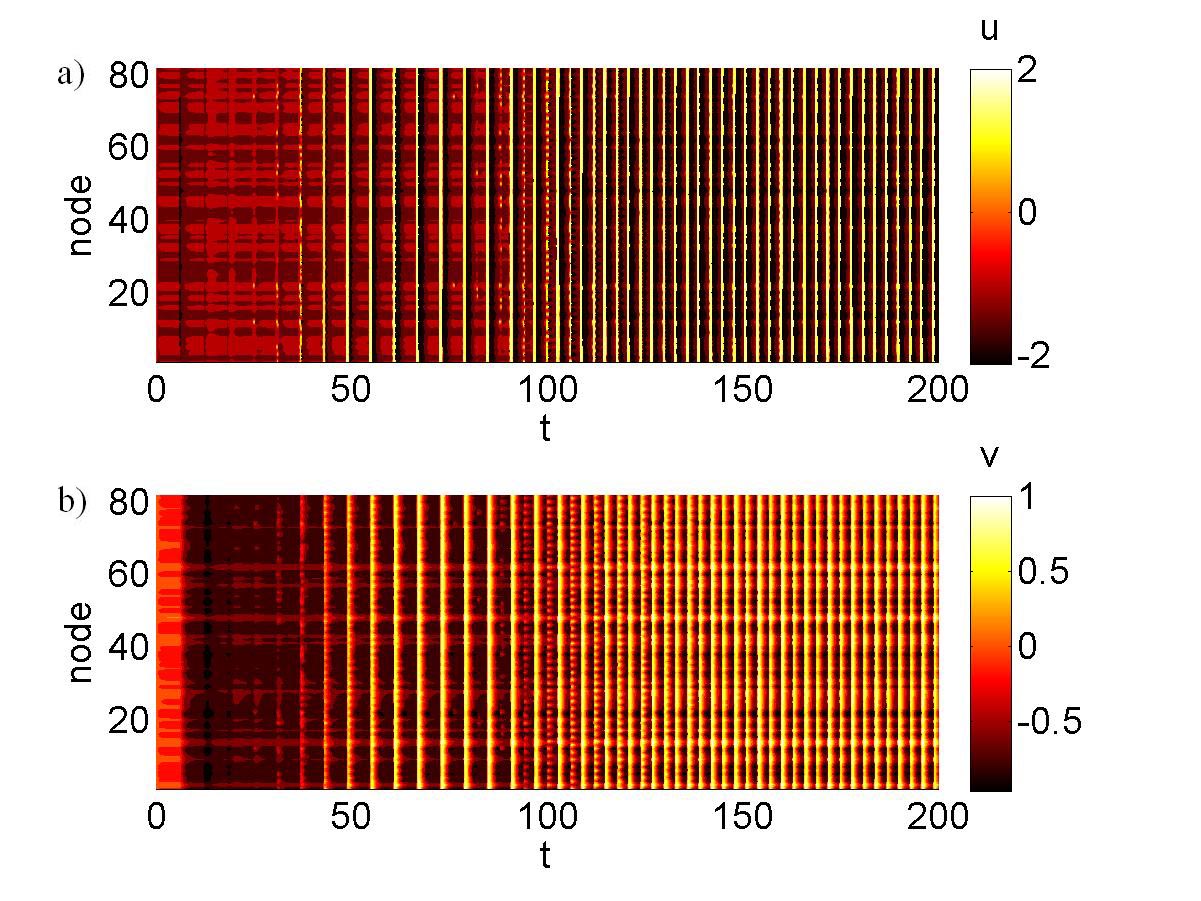}}
\end{minipage}
\caption{Dynamics of 82 FitzHugh-Nagumo
  systems given by Eq.~\eqref{m} with Cantor-type hierarchical topology. (a) and (b): time series of the activator
  and the inhibitor of all nodes, respectively. Parameters:  $C=0.065$, $\tau=6$.  Other parameters and initial conditions as in Fig.~\ref{figC}.}
\label{figTau}
\end{figure}

\begin{figure}
\begin{minipage}[h]{1\linewidth}
\center{\includegraphics[width=1\linewidth]{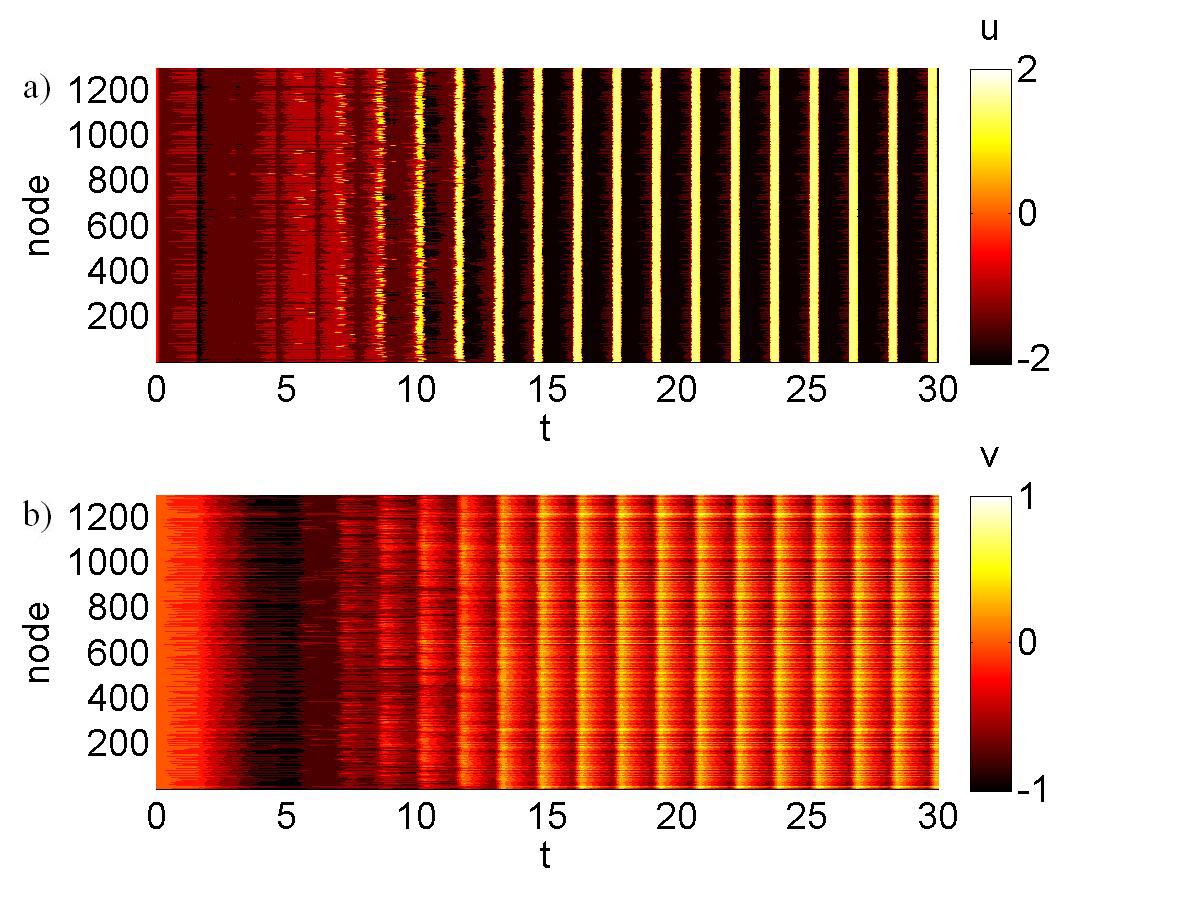}}
\end{minipage}
\caption{Dynamics of 1297 FitzHugh-Nagumo
  systems given by Eq.~\eqref{m} with Cantor-type hierarchical topology. (a) and (b): time series of the activator
  and the inhibitor of all nodes, respectively. Parameters:
  $B=[1~0~1~0~0~0]$, $b=6$, $n=4$, $N=1297$, $C=0.04$. Other parameters and
  initial conditions as in Fig.~\ref{figC}.}
\label{figTop}
\end{figure}

\begin{figure}
\begin{minipage}[h]{1\linewidth}
\center{\includegraphics[width=1\linewidth]{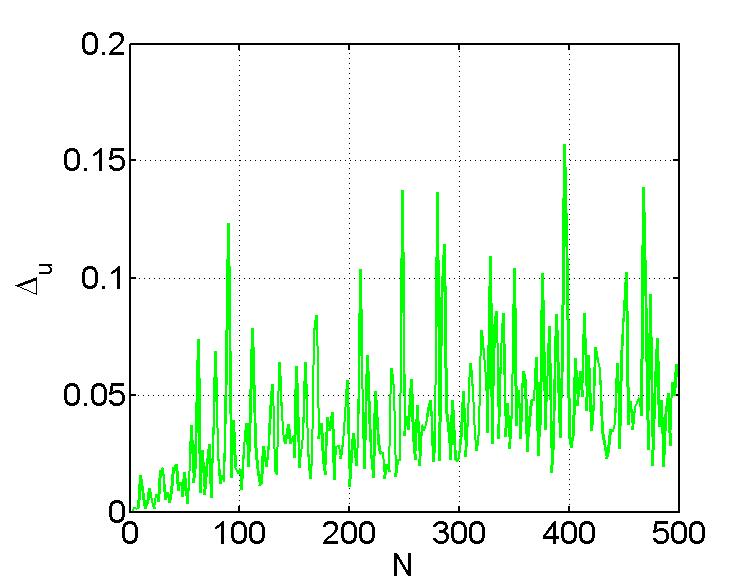}}
\end{minipage}
\caption{Dependence of the activator synchronization index $\Delta_u$
  on the number of nodes $N$ for the FitzHugh-Nagumo network given by
  Eq.~\eqref{m} with bidirectional ring topology and local coupling, Eq.~\eqref{adj}. Parameters: $\tau=1.5$, $\epsilon=0.01$, $C=1$. The
  threshold parameters $a_i$ are normally distributed with mean $\mu=1$ and standard deviation $\sigma=0.1$. Initial conditions: $u_i(t)=0$, $v_i(t)=0$, $i=1,\dots,N$,  for $t\in[-\tau,0]$.}
\label{figN}
\end{figure}

We now study the influence of the delay $\tau$ on the dynamics of
the network. To this end, we fix the coupling strength at $C=0.065$
and  the standard deviation at $\sigma=0.1$. For these parameters we
always obtain synchronization, however, the delay affects the transient time.  For example, for $\tau=6$ the transient
time takes approximately $110$ units of time (see
Fig.~\ref{figTau}). Figure~\ref{figTau}a shows that during the
transient time, the network already synchronizes but with a lower
frequency than the final one. 
Note that many authors have emphasized the
importance of the transient phenomena that arise in neural networks
with delay \cite{GRO99,MIL10,MIL12,PAK98,QUA11}. However, for the
chosen parameters we have not observed these phenomena.

Next we consider how synchronization depends on the base pattern $B$. A systematic study of the influence of different base patterns has been performed for networks of identical Van der Pol oscillators in the context of chimera states \cite{ULO16}. In agreement
with the synchronization condition  $C>1/n_\text{min}$ (cf. Sec.~\ref{Sec:synch_analys}), the numerical simulations show that synchronization
depends only on the number of elements equal to $1$ in each row
of the connectivity matrix $\mathbf{G}$. We carry out simulations with
different bases of $b=6$ elements with $n=4$ iteration and $c_1=2$
nonzero entries, and obtain synchronization in all cases for
$C=0.065$. As an example we show in Fig.~\ref{figTop} the results of simulations
of the  FHN network with $C=0.04$ and the base pattern $B=[1~0~1~0~0~0]$. Note that
for $C=0.065$ and base pattern $B=[1~0~1~0~0~0]$ there is also synchronization
(not shown here), however, the network is in the excitable regime, because
the coupling strength is too high for the  network to be in the oscillatory regime as discussed in Sec.~\ref{Sec:coupl_strength_analys}.

Finally, we investigate how the size of the network influences its
synchronizability. Since in the case of hierarchical
topologies the number of nodes $N$ is given by $b^n+1$ where $b$ is the length
of the base and $n$ the number of iterations, the hierarchical network
does not allow for increasing $N$ in equally-sized small
steps. Therefore, we choose here a bidirectionally locally coupled ring  given
by the following adjacency matrix
\begin{equation}\label{adj}
\mathbf{G} = \begin{pmatrix}
0 & 1 & 0 &\cdots & 1 \\
1 & 0 & 1 &\cdots & 0 \\        
\vdots & \vdots & \vdots &\ddots & \vdots \\
0 & 0 & 0 &\cdots & 1 \\
1 & 0 & 0 &\cdots & 0
\end{pmatrix}.
\end{equation}
Thus, each node has only two (bidirectional) connections for any size of the system.
We fix the coupling strength at $C=1$ and  the standard deviation at
$\sigma=0.1$. For $C=1>1/n_\text{min}=0.5$ we expect synchronization (cf. Sec.~\ref{Sec:synch_analys}).
The results of the simulation are presented in
Fig.~\ref{figN}, where the synchronization index vs. the number of
nodes is shown. One can see that $\Delta_u<2\sigma$, $\forall N$. This means that the network
is synchronized with precision $2\sigma$ for all $N$. We conclude that
synchronization does not depend on the network size but only on the
coupling strength and the number of connections to each node of the
network. Note, however, that the quality of synchronization, i.e., the
value of $\Delta_u$ slightly depends on the network size.

\section{Control of synchronization in FitzHugh-Nagumo network}\label{Sec:control}
In this section we study how to control synchronization
in the case where the synchronization condition $C>1/n_\text{min}$  is not
met. For this purpose a  mean field control, i.e., adding the term
$I=\frac{1}{N}\sum_{i=1}^{N}u_i$ to each node, has been considered, for
example in Refs.~\onlinecite{ROS04,ROS04a}. 
However, here we suggest a different approach. We will use
the control $I(t)$ in  the form
\begin{equation}\label{f8}
I(t)=\gamma u(t),
\end{equation}
where $\gamma$ is a control gain and $u$ is an activator value of the \textit{master} system, which is described by the following equations
\begin{equation}\label{f9}
\begin{aligned}
\epsilon \dot u&=u-\frac{u^3}{3}-v,\\
\dot v&=u+a \\
\end{aligned}
\end{equation}
where $v$ is an inhibitor value and $a$ is a desired threshold.
Compared to a mean field control this approach has the advantage that
it is not necessary to measure and average some output variable.
Furthermore, even if all oscillators are in the excitable regime the
control ensures synchronization in an  oscillatory state if we choose
$|a|<1$ in Eq.~\eqref{f9}.

In the following, we will use $\gamma=0.3$ and $a=0.9$, i.e.,the
\textit{master} system is in the oscillatory regime.
We will demonstrate the control in the form Eq.~\eqref{f8} on different network topologies. 

\subsection{Bidirectional ring topology}\label{Sec:ring_net_control}

\begin{figure}
\begin{minipage}[h]{1\linewidth}
\center{\includegraphics[width=1\linewidth]{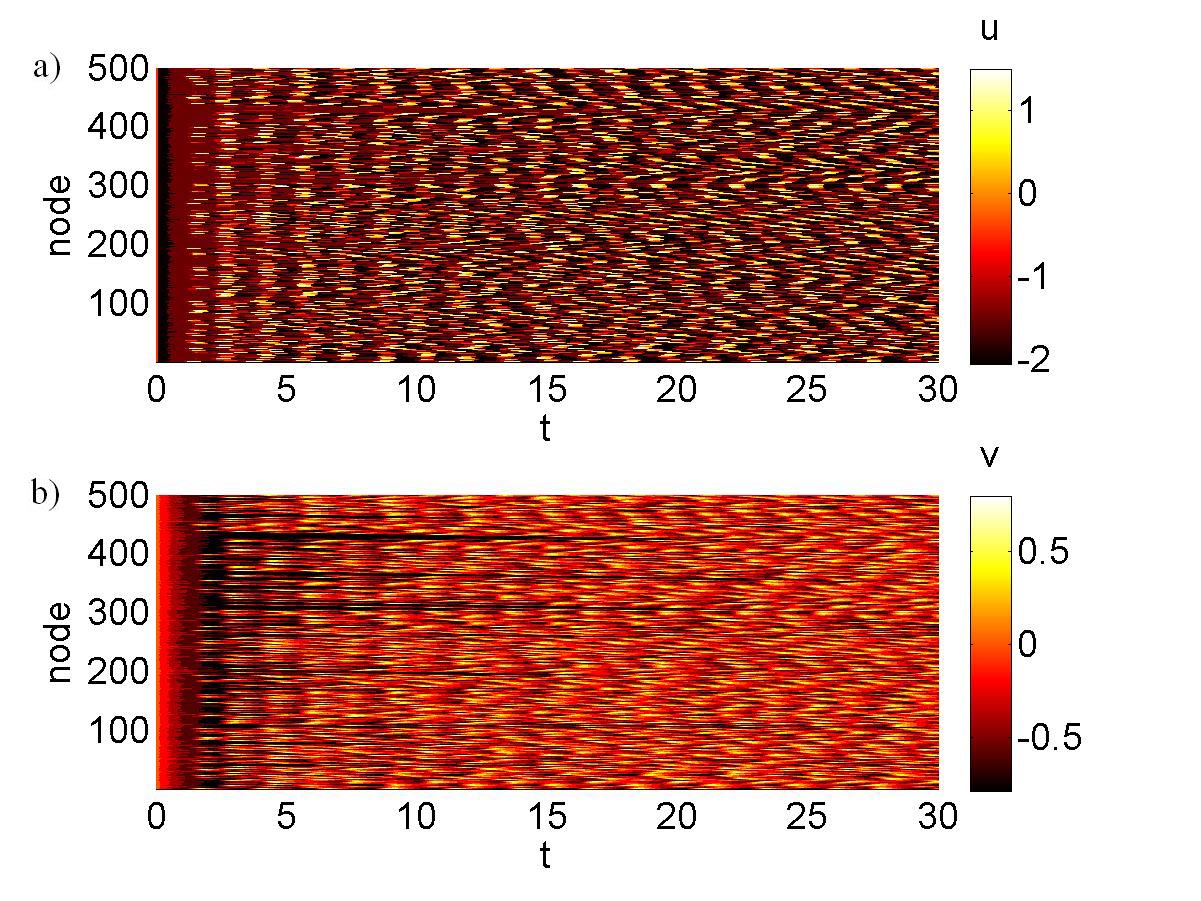}}
\end{minipage}
\caption{Dynamics of 500 FitzHugh-Nagumo
  systems given by Eq.~\eqref{m} with a locally coupled ring topology Eq.~\eqref{adj} without control, i.e., $I(t)=0$. (a) and (b): time series of the activator
  and the inhibitor of all nodes, respectively. Parameters: $N=500$,
  $C=0.1$, $\tau=1.5$, $\epsilon=0.01$. The threshold parameters $a_i$
  are normally distributed with mean $\mu=1$ and standard deviation $\sigma=0.1$, truncated at $a_i=\mu\pm\sigma$. Initial conditions: $u_i(t)=0$, $v_i(t)=0$, $i=1,\dots,N$,  for $t\in[-\tau,0]$.}
\label{fig2}
\end{figure}

\begin{figure}
\begin{minipage}[h]{1\linewidth}
\center{\includegraphics[width=1\linewidth]{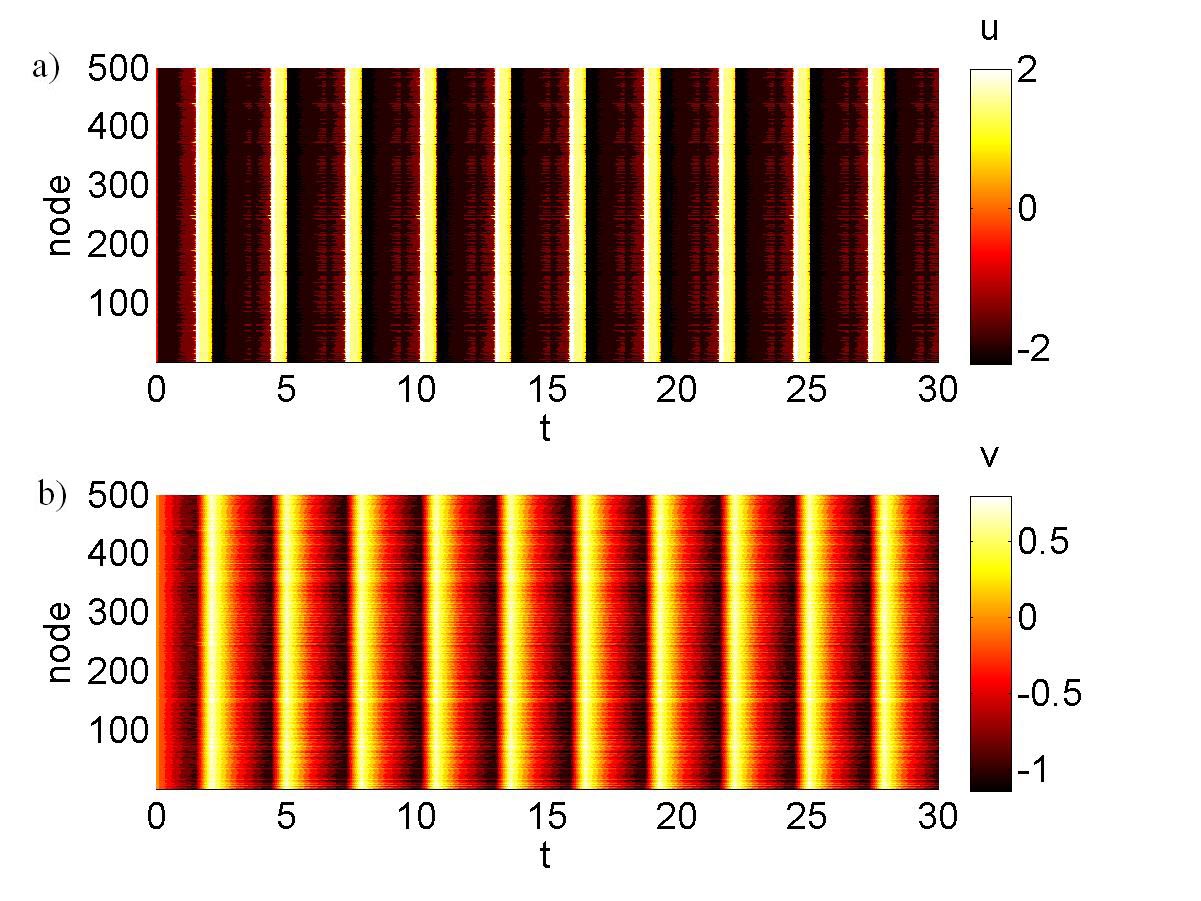}}
\end{minipage}
\caption{Control of synchronization in 500 FitzHugh-Nagumo
  systems with a locally coupled ring topology Eq.~\eqref{adj} by the external stimulus
  $I(t)$ according to Eqs.~\eqref{f8} and \eqref{f9}. (a) and (b): time series of the activator and the inhibitor of all nodes, respectively. Parameters: $\gamma=0.3$, $a=0.9$. Other parameters and initial conditions as in Fig. \ref{fig2}.}
\label{fig3}
\end{figure}

We start our consideration with a one-dimensional ring of $N$
delay-coupled FHN oscillators, where each element is coupled to its
neighbors on both sides, i.e., the system \eqref{m} with the connectivity matrix
$\mathbf{G}$ given by Eq.~~\eqref{adj}.

For the simulation we consider the case of $N=500$ nodes. We use
a coupling strength $C$ equal to $0.1$. Without control the network
does not synchronize as can be seen in  Fig.~\ref{fig2}, where the
simulation of a network without control, i.e., $I(t)=0$, is shown.
Clearly, there is no synchronization between the activator and inhibitor values. 

To synchronize this network we use the control in the form of
Eqs.~\eqref{f8} and \eqref{f9}. The result is shown in 
Figure~\ref{fig3}:  The
 control goal is achieved and synchronization holds for the activators (Fig.~\ref{fig3}a) and the inhibitors (see Fig.~\ref{fig3}b).

\subsection{Hierarchical network topology}\label{Sec:hier_net_control}

\begin{figure}
\begin{minipage}[h]{1\linewidth}
\center{\includegraphics[width=1\linewidth]{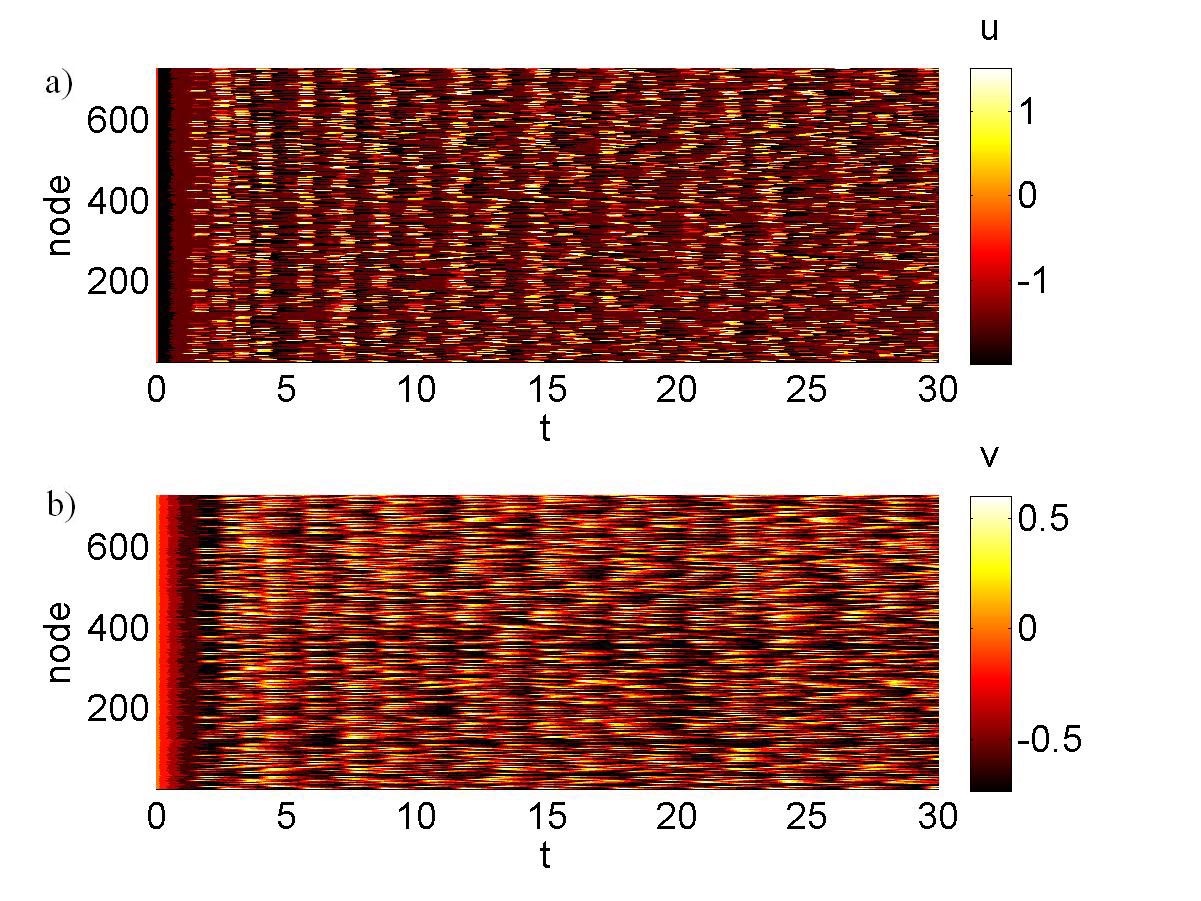}}
\end{minipage}
\caption{Dynamics of 730 FitzHugh-Nagumo
  systems with Cantor-type hierarchical topology without control, i.e., $I(t)=0$. (a) and (b): time series of the activator
  and the inhibitor of all nodes, respectively. Parameters: $n=6$, $N=730$, $C=0.001$. Other parameters and initial conditions as in Fig.~\ref{figC}.}
\label{fig4}
\end{figure}

\begin{figure}
\begin{minipage}[h]{1\linewidth}
\center{\includegraphics[width=1\linewidth]{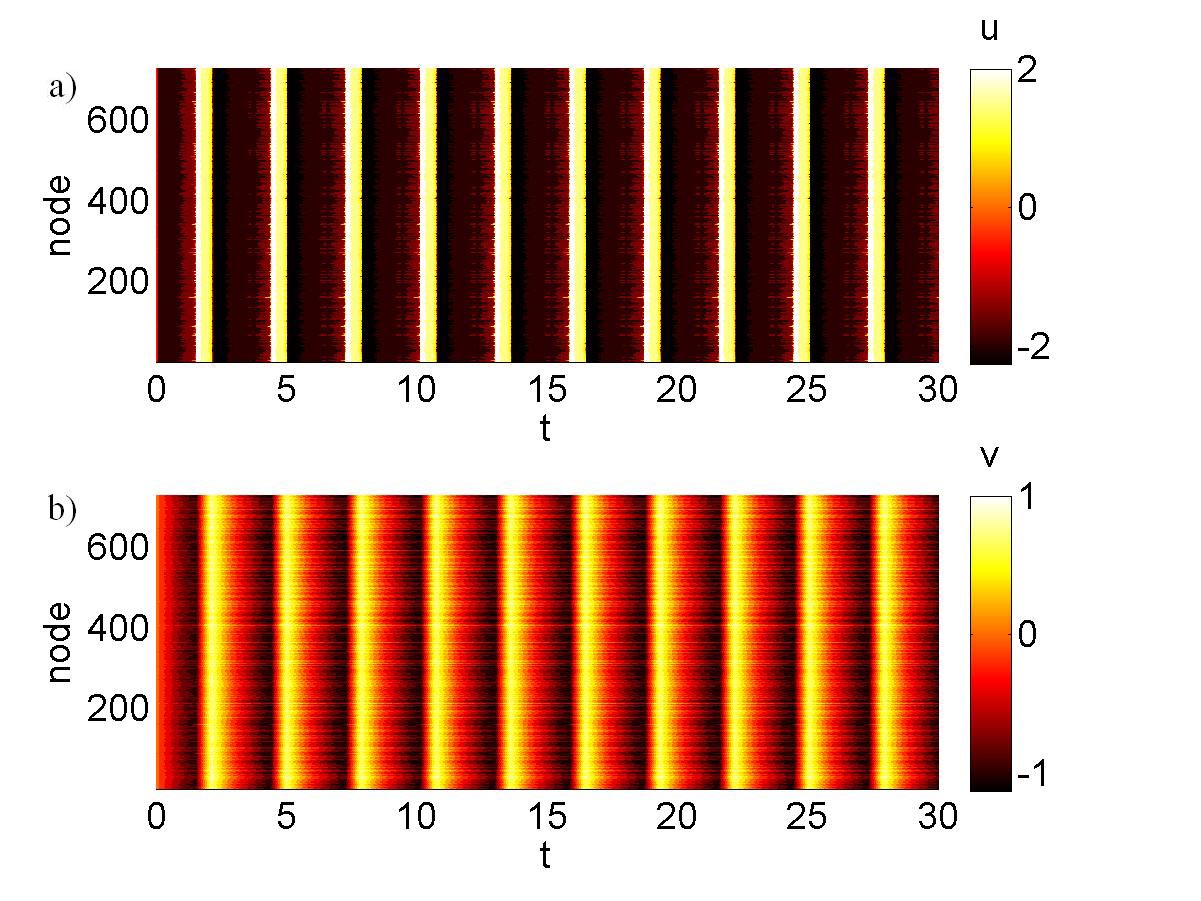}}
\end{minipage}
\caption{Control of synchronization in 730 FitzHugh-Nagumo
  systems with Cantor-type hierarchical topology by external stimulus $I(t)$
  according to Eqs.~\eqref{f8} and \eqref{f9}. (a) and (b): time series of the activator
  and the inhibitor of all nodes, respectively. Parameters: $\gamma=0.3$, $a=0.9$. Other parameters and initial conditions as in Fig. \ref{fig4}.}
\label{fig5}
\end{figure}

Next we investigate a hierarchical topology with base pattern $B=[1~0~1]$ and
iteration number $n=6$, i.e., the coupling matrix $\mathbf{G}$ has the
dimension $N\times N$ with $N=3^6+1=730$ (for the construction of a
hierarchical network
see Sec.~\ref{Sec:analys_hier}). We use a very weak coupling strength
$C$ equal to $0.001$. Figure~\ref{fig4} shows the results of the
simulation of network behavior without control, i.e., $I(t)=0$. Clearly,
 no synchronization between the nodes takes place.

To synchronize this network we use again the control $I(t)$ given by
Eqs.~\eqref{f8} and \eqref{f9}.  Figure~\ref{fig5} presents the results of a
 simulation of the FHN network \eqref{m} with hierarchical topology
 and the external stimulus $I(t)$. The
 control goal is achieved and synchronization holds (see the time
 series of the activators and inhibitors in Fig.~\ref{fig5}(a), and (b), respectively). 
 
\subsection{Random network topology}\label{Sec:rand_net_control}

\begin{figure}
\begin{minipage}[h]{1\linewidth}
\center{\includegraphics[width=1\linewidth]{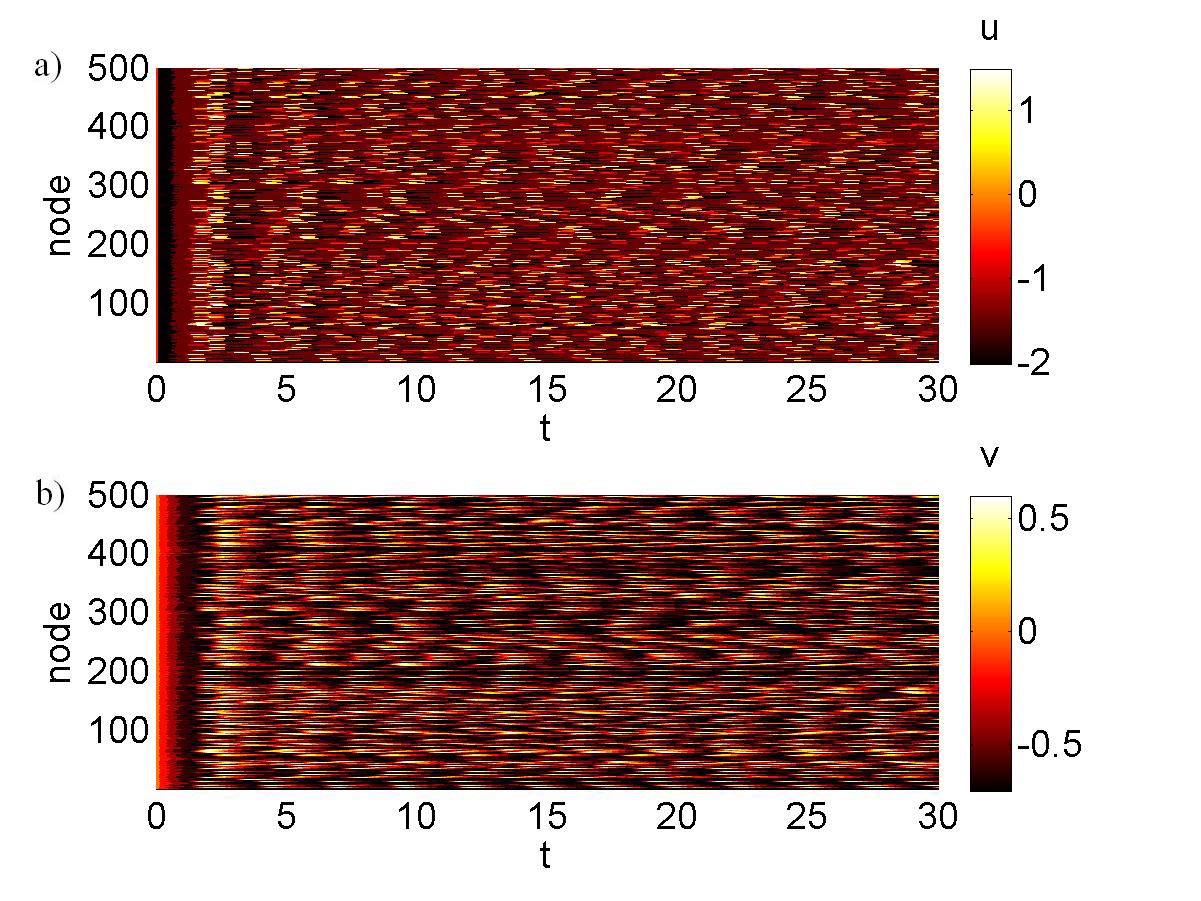}}
\end{minipage}
\caption{Dynamics of 500 FitzHugh-Nagumo
  systems with a random topology without control, i.e., $I(t)=0$. (a) and (b): time series of the activator
  and the inhibitor of all nodes, respectively. Parameters: $N=500$. The nonzero entries are drawn from a Gaussian distribution with mean
$\mu_C=1$ and variance $\sigma_C^2=1$, truncated at $a_i=\mu\pm\sigma$. Parameters: $C=0.0001$. Other parameters and initial conditions as in Fig. \ref{fig2}.}
\label{fig6}
\end{figure}

\begin{figure}
\begin{minipage}[h]{1\linewidth}
\center{\includegraphics[width=1\linewidth]{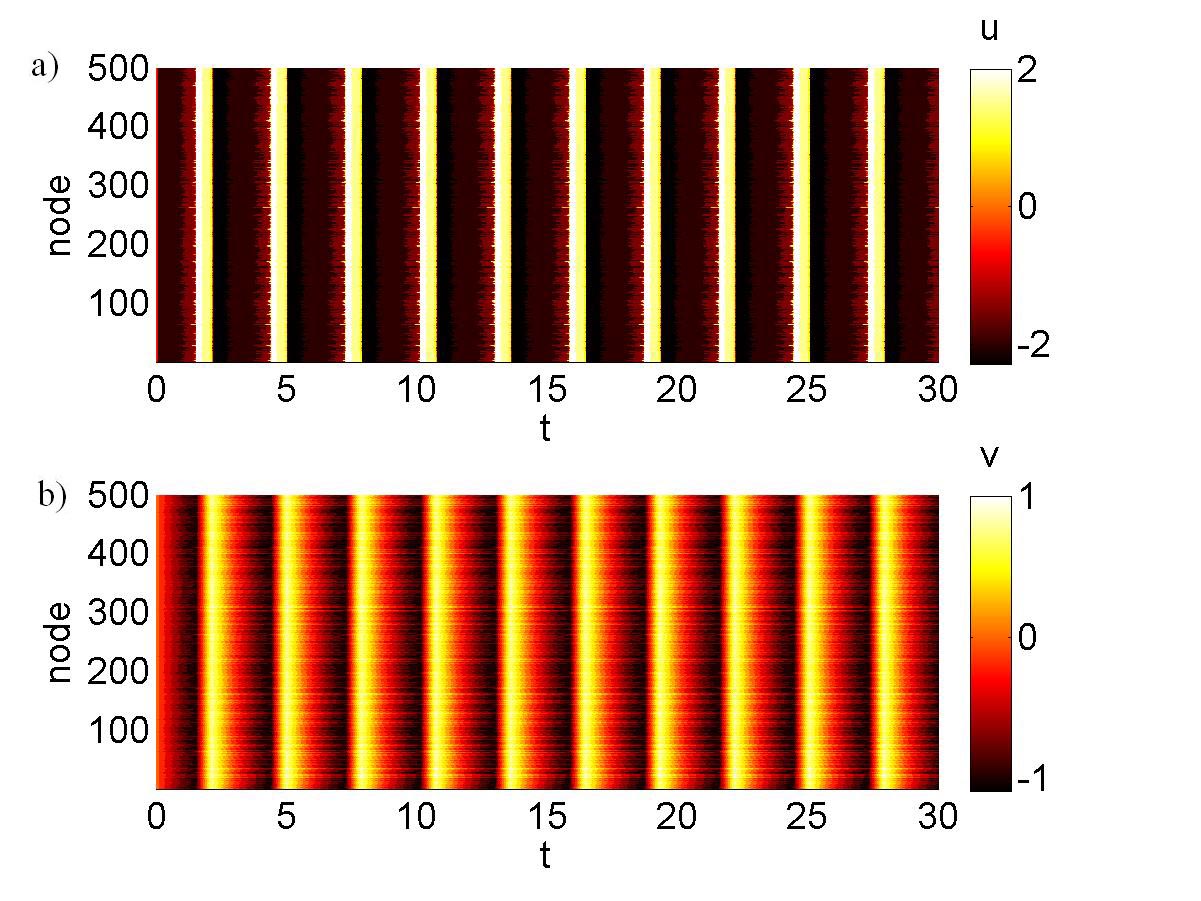}}
\end{minipage}
\caption{Control of synchronization in 500 FitzHugh-Nagumo
  systems with a random topology by external stimulus $I(t)$ according
  to Eqs.~\eqref{f8} and \eqref{f9}. (a) and (b): time series of the activator
  and the inhibitor of all nodes, respectively. Parameters: $\gamma=0.3$, $a=0.9$. Other parameters and initial conditions as in Fig. \ref{fig6}}
\label{fig7}
\end{figure}

Last we consider a network of $N=500$ nodes with random topology. We
use a sparse matrix with link density equal to $0.3$. This means there are
approximately $0.3N^2$ normally distributed nonzero entries.  These
nonzero entries are drawn from a Gaussian distribution with mean
$\mu_C=1$ and variance $\sigma_C^2=1$, truncated at $a_i=\mu\pm\sigma$. This process results in a
weighted random network. Note that
in this case the average node degree is much higher compared to the
previously studied topologies. Thus, only for very weak
coupling strengths we anticipate  non-synchronous dynamics (see
Sec.~\ref{Sec:synch_analys}), in which case the need of control 
arises.

In fact, for $C=0.0001$ no synchronization takes place as can be seen
in Fig.~\eqref{fig6}. Applying the control in the form of
Eqs.~\eqref{f8} and \eqref{f9} synchronizes the network as  shown for
 the activators and inhibitors in Figs.~\ref{fig7}(a) and (b), respectively. 

\section{Conclusion}\label{Sec:concl}
We have considered synchronization and its control in networks of heterogeneous FitzHugh-Nagumo systems, a
neural model which is considered to be generic for excitable
systems close to a Hopf bifurcation. To this end, we have investigated
networks with heterogeneous threshold parameters.
It is well known that networks with
heterogeneous nodes are much less likely to synchronize than networks of
identical nodes. Furthermore, synchrony will take place in a state
where the trajectories of the different nodes are not identical but
small deviations can be observed. To counteract this effect of
heterogeneity we have proposed an algorithm for controlling synchrony.

First,  we have generalized the analytic conditions for the
occurrence of the Hopf bifurcations obtained in
Refs.~\onlinecite{SCH08,PLO15} for unidirectional locally coupled ring networks. In this way, we have been able to determine the threshold values for which
the system is in the excitable regime.
Afterwards, we have considered the case that all but one node are in
synchrony and derived a critical coupling strength depending on the minimum node degree of the network.
Above the critical coupling strength the node will synchronize with
the rest of the network. However, if the coupling strength becomes too
large, synchronization still holds but amplitude death sets in. Thus,
synchronization in an oscillatory state is only possible for
intermediate coupling strengths. We have studied synchronization in
networks with Cantor-type hierarchical topologies. The study of different
hierarchical architectures in the neuron connectivity is motivated by
MRI results of the brain structure which shows that the neuron axon network spans the brain area fractally and not homogeneously. The results of these numerical investigations match with our analytical synchronization condition.

Based on the result that the network does not synchronize for too small
coupling strengths, we have suggested to add the same
external stimulus to all nodes to ensure synchronization in networks
where it is absent without control. We have considered the activator value of a master neuron as an external stimulus. 
We have then applied the control algorithm to different types of networks
(bidirectional ring networks, hierarchical networks, random
networks), and the simulation results have shown synchronization. Since we use the activator value of the neuron in the oscillatory
state, we suppose that it is possible to consider other periodic
functions with appropriate frequency as a control to ensure the
synchronization, and we have confirmed this in simulations (not
shown in the paper). Furthermore, we suggest that our control algorithm
and obtianed synchronization conditions can also help with study of desynchronization of the network in cases where this is desirable, for
example in the case of abnormal synchrony in the brain as in epilepsy
or Parkinson disease.
 Given the paradigmatic nature of the FitzHugh-Nagumo system, we expect our investigations to be applicable in a wide range of excitable systems, for instance, to the other models for neural spiking \cite{WIL99,IZH03,NAU08}.

\begin{acknowledgements} 
           
This work is supported by the German-Russian Interdisciplinary
Science Center (G-RISC) funded by the German Federal Foreign Office
via the German Academic Exchange Service (DAAD). J.L. and E.S.
acknowledge support by Deutsche Forschungsgemeinschaft (DFG) in the
framework of SFB 910. S.P. and A.F. acknowledge support by SPbSU (Grant No. 6.38.230.2015) and by the Goverment of Russian Federation (Grant  No. 074-U01). Section~\ref{Sec:FHN_analys} was performed in IPME RAS, supported solely by RSF (Grant No. 14-29-00142).
\end{acknowledgements}


\begin{thebibliography}{20}

\bibitem{BLE88}
I.I. Blekhman, {\em Synchronization in Science and Technology} (ASME Press, 1988).

\bibitem{PIK01}
A. Pikovsky, M. Rosenblum, and J. Kurths, {\em Synchronization: A
  universal concept in nonlinear sciences} (Cambridge University Press,
 Cambridge, 2001).

\bibitem{WIE95}
K. Wiesenfeld, and J.W. Swift, Phys. Rev. E {\bf 51}, 1020 (1995).

\bibitem{WIE90}
K. Wiesenfeld {\em et al.}, Phys. Rev. Lett. {\bf 65}, 1749 (1990).

\bibitem{PES75}
C.S. Peskin, {\em Mathematical Aspects of Heart Physiology} (Courant Institute of Mathematical Sciences, New York, 1975).

\bibitem{BUC68}
J. Buck and E. Buck, Science {\bf 159}, 1319 (1968).

\bibitem{WAL69}
T.J. Walker, Science {\bf 166}, 891 (1969).

\bibitem{KIS02}
I. Kiss, Y. Zhai, and J. Hudson, Phys. Rev. Lett. {\bf 88}, 238301 (2002);
Science {\bf 296}, 1676 (2002).

\bibitem{TAS99}
P.A. Tass, {\em Phase Resetting in Medicine and Biology: Stochastic Modelling and Data Analysis} (Springer, Berlin, 1999).

\bibitem{GOL01}
D. Golomb, D. Hansel, and G. Mato, in {\em Neuro-informatics and Neural Modeling}, Handbook of Biological Physics Vol. 4, edited by F. Moss and S. Gielen (Elsevier, Amsterdam, 2001), pp. 887-968.

\bibitem{MIL03}
{\em Epilepsy as a Dynamic Disease}, edited by J. Milton and P. Jung (Springer, Berlin, 2003).

\bibitem{CHK78}
S.A. Chkhenkeli, Bull. Georgian Acad. Sci. {\bf 90}, 406 (1978).

\bibitem{BEH91}
A. Behabid {\em et al.}, Lancet {\bf 337}, 403 (1991).


  

  

\bibitem{ZHO08}
J. Zhou, J. Lu, and J. L\"u, Automatica {\bf 44},  996 (2008).

\bibitem{LU09}
X. Lu, and B. Qin, Phys. Lett. A {\bf 373},  3650 (2009).

\bibitem{LU12}
J. Lu, J. Kurths, J. Cao, N. Mahdavi, and C. Huang, IEEE Trans. Neural Netw. Learn. Syst. {\bf 23},  285 (2012).

\bibitem{SEL12}
A.A. Selivanov, J. Lehnert, T. Dahms, P. H{\"o}vel, A.L. Fradkov, and E. Sch{\"o}ll, Phys. Rev.~E {\bf 85},
  016201 (2012).
  
 \bibitem{GUZ13}
P.Y. Guzenko, J. Lehnert, and E. Sch{\"o}ll, Cybernetics and Physics {\bf 2},  15 (2013).

\bibitem{LEH14}
J. Lehnert, P. H{\"o}vel, A.A. Selivanov, A.L. Fradkov, and E. Sch{\"o}ll Phys.~Rev.~E {\bf 90},  042914 (2014).

\bibitem{SCH16}
E. Sch{\"o}ll, S.H.L. Klapp, and P. H{\"o}vel, {\em Control of Self-Organizing Nonlinear Systems},  (Springer, Berlin, 2016).

\bibitem{STR00}
S.H. Strogatz, Physica D {\bf 143}, 1 (2000).

\bibitem{SUN09a}
J. Sun, E.M. Bollt, and T. Nishikawa, Europhys. Lett. {\bf 85},  60011 (2009).

\bibitem{ROS04}
M.G. Rosenblum, A. Pikovsky, Phys.~Rev.~E {\bf 70}, 041904 (2004).

\bibitem{ROS04a}
M.G. Rosenblum, A. Pikovsky, Phys.~Rev.~Lett. {\bf 92}, 114102 (2004).

\bibitem{POP05}
O.V. Popovych,C. Hauptmann, P.A. Tass, Phys. Rev. Lett. {\bf 94}, 164102 (2005).

\bibitem{FIT61a}
R. FitzHugh, Biophys. J. {\bf 1},  445  (1961).

\bibitem{NAG62}
J. Nagumo, S. Arimoto, and S. Yoshizawa., Proc. IRE {\bf 50},  2061  (1962).

\bibitem{VUK14}
V. Vuksanovi\'c, and P. H{\"o}vel, NeuroImage {\bf 97}, 1 (2014).

\bibitem{KAT09}
P. Katsaloulis, D. A. Verganelakis, and A. Provata, Fractals {\bf 17}, 181 (2009).

\bibitem{EXP11}
P. Expert, T.S. Evans, V. D. Blondel, and R. Lambiotte, Proc. Natl. Acad. Sci. U.S.A. {\bf 108}, 7663 (2011).

\bibitem{PRO12}
A. Provata, P. Katsaloulis, and D.A Verganelakis, Chaos Soliton. Fract. {\bf 45}, 174 (2012).

\bibitem{OME15}
I. Omelchenko, A. Provata, J. Hizanidis, E. Sch{\"o}ll, and P. H{\"o}vel, Phys.~Rev.~E {\bf 91}, 022917 (2015).




  

\bibitem{LIN04}
B. Lindner, J. Garc{\'i}a-Ojalvo, A.B. Neiman, and L. Schimansky-Geier,
  Phys.~Rep. {\bf 392},  321  (2004).
  
  \bibitem{HEI10}
M. Heinrich, T. Dahms, V. Flunkert, S.W. Teitsworth, and E. Sch{\"o}ll,
  New~J.~Phys. {\bf 12},  113030  (2010).
  
  \bibitem{NAS04}
 M. P. Nash, A. V. Panfilov,  Prog. Biophys. Mol. Biol. {\bf 85}, 501 (2004). 

\bibitem{WIN91}
A. T. Winfree, Chaos  {\bf 1}, 303 (1991). 

\bibitem{GAN02}
A. Ganopolski, S. Rahmstorf, , Phys. Rev. Lett.  {\bf 88},  038501 (2002)

\bibitem{MUR93}
J.D. Murray, {\em Mathematical Biology}, Vol.~19 of {\em Biomathematics
  Texts}, 2nd ed. (Springer, Berlin Heidelberg, 1993).

\bibitem{IZH00a}
E.M. Izhikevich, Int.~J.~Bifur.~Chaos {\bf 10},  1171  (2000).

\bibitem{DAH08c}
M.A. Dahlem, G. Hiller, A. Panchuk and E. Sch{\"o}ll, Int.~J.~Bifur.~Chaos \textbf{19},  745 (2009).

\bibitem{SCH08}
 E. Sch{\"o}ll, G. Hiller, P. H{\"o}vel, and M.A. Dahlem, Phil. Trans.~R. Soc.~A. {\bf 367}, 1079 (2009).

\bibitem{PLO15}
S.A. Plotnikov, J. Lehnert, A.L. Fradkov, and E. Sch{\"o}ll, Int.~J.~Bifur.~Chaos {\bf 26}, 1650058 (2016).





\bibitem{KAT12}
P. Katsaloulis, A. Ghosh, A. C. Philippe, A. Provata, and R. Deriche, Eur. Phys. H. B {\bf 85}, 1 (2012).

\bibitem{MAN83}
B. B. Mandelbrot, {\em The Fractal Geometry of Nature}, 3rd ed. (W. H. Freeman and Comp., New York, 1983).

\bibitem{FED88}
J. Feder, {\em Fractals} (Plenum Press, New York, 1988).

\bibitem{GRO99}
C. Grotta-Ragazzo, K. Pakdaman, and C.P. Malta, Phys. Rev. E {\bf 60}, 6230 (1999).

\bibitem{MIL10}
J. Milton, P. Naik, C. Chan, and S.A. Campbell, Math. Model. Nat. Phenom. {\bf 5}, 125 (2010).

 \bibitem{MIL12}
J.G. Milton, Eur. J. Neurosci. {\bf 36}, 2156 (2012).

 \bibitem{PAK98}
 K. Pakdaman, C. Grotta-Ragazzo, C.P. Malta, O. Arino, and J.-F. Vobert, Neural Net. {\bf 11}, 509 (1998).
 
 \bibitem{QUA11}
A. Quan, I. Osorio, T. Ohira, and J. Milton, Chaos {\bf 21}, 047512 (2011).

\bibitem{ULO16}
S. Ulonska, I. Omelchenko, A. Zakharova, and E. Sch{\"o}ll, arXiv:1603.00171 (2016).

\bibitem{WIL99}
H.R. Wilson, J. Theor. Biol. {\bf 200}, 375 (1999).

\bibitem{IZH03}
E.M. Izhikevich, IEEE Trans. Neural Net. {\bf 14}, 1569 (2003).

\bibitem{NAU08}
R. Naud, N. Marcille, C. Clopath, W. Gerstner, Biol. Cybern. {\bf 99}, 335 (2008).







\end{thebibliography}
\end{document}